\documentstyle[12pt,equation,double]{article}
\begin{document}
\setlength{\baselineskip}{15pt}
\newcommand{\aq}{\mbox{$a_{\tilde{Q}}$}}
\newcommand{\bq}{\mbox{$b_{\tilde{Q}}$}}
\newcommand{\asq}{\mbox{$a^2_{\tilde{Q}}$}}
\newcommand{\bsq}{\mbox{$b^2_{\tilde{Q}}$}}
\newcommand{\tanb}{\mbox{$\tan \! \beta$}}
\newcommand{\mlsq}{\mbox{$m^2_{H_2}$}}
\newcommand{\mhsq}{\mbox{$m^2_{H_1}$}}
\newcommand{\mx}{\mbox{$M_X$}}
\newcommand{\rt}{\mbox{$\sqrt{|\Delta|}$}}
\newcommand{\mc}{\mbox{$m_{\chi}$}}
\newcommand{\mcsq}{\mbox{$m^2_{\chi}$}}
\newcommand{\msq}{\mbox{$m_{\tilde{q}}$}}
\newcommand{\msqsq}{\mbox{$m^2_{\tilde{q}}$}}
\newcommand{\msqiv}{\mbox{$m^{-2}_{\tilde{q}}$}}
\newcommand{\msqf}{\mbox{$m^{-4}_{\tilde{q}}$}}
\newcommand{\msQ}{\mbox{$m_{\tilde{Q}}$}}
\newcommand{\mQsq}{\mbox{$m^2_Q$}}
\newcommand{\mqsq}{\mbox{$m^2_q$}}
\newcommand{\msQsq}{\mbox{$m^2_{\tilde{Q}}$}}
\newcommand{\msbsq}{\mbox{$m^2_{\tilde{b}_1}$}}
\newcommand{\msb}{\mbox{$m_{\tilde{b}_1}$}}
\newcommand{\mstau}{\mbox{$m_{\tilde{\tau}_1}$}}
\newcommand{\mst}{\mbox{$m_{\tilde{t}_1}$}}
\newcommand{\mstsq}{\mbox{$m^2_{\tilde{t}_1}$}}
\newcommand{\gsq}{\frac {g_s^2}{16 \pi^2}}
\newcommand{\be}{\begin{equation}}
\newcommand{\ee}{\end{equation}}
\newcommand{\een}{\end{subequations}}
\newcommand{\ben}{\begin{subequations}}
\newcommand{\beq}{\begin{eqalignno}}
\newcommand{\eeq}{\end{eqalignno}}
\renewcommand{\thefootnote}{\fnsymbol{footnote} }
\noindent
\begin{flushright}
MAD/PH/768\\
June 1993
\end{flushright}
\vspace{1.5cm}
\pagestyle{empty}
\begin{center}
{\Large \bf Neutralino--Nucleon Scattering Revisited}\\
\vspace{5mm}
Manuel Drees\footnote{Heisenberg Fellow} and
Mihoko M. Nojiri\footnote{Address after July 1, 1993: Theory group, KEK,
Tsukuba, Japan. E--mail: NOJIRI@WISCPHEN}\\
{\em Physics Department, University of Wisconsin, Madison, WI 53706, USA}
\end{center}

\begin{abstract}

We present a detailed discussion of the elastic scattering of a supersymmetric
neutralino off a nucleon or nucleus, with emphasis on the spin--independent
interaction. We carefully treat QCD effects on the squark exchange
contribution. In particular, we identify a class of terms that survive even in
the absence of mixing in both the neutralino and squark sectors; the
corresponding quark and gluon operators also appear in analyses of
deep--inelastic lepton--nucleon scattering (``twist--2 operators''), so their
hadronic matrix elements are well known. We also re--emphasize the importance
of mixing between the superpartners of left-- and right--handed quarks, and of
the contribution from the heavier scalar Higgs boson. We use our refined
calculation of the scattering amplitude to update predictions of signal rates
for cosmic relic neutralino searches with Germanium detectors. In general the
counting rate varies strongly with the values (even the signs) of model
parameters; typical results lie between a few times $10^{-4}$ and a few times
$10^{-1}$ events/(kg$\cdot$day).

\end{abstract}

\clearpage
\noindent
\setcounter{footnote}{0}
\pagestyle{plain}
\setcounter{page}{1}
\section*{1) Introduction}
The lightest supersymmetric particle (LSP) is one of the most attractive
candidates for the dark matter (DM) in the Universe \cite{1}.
It is theoretically well motivated, since all supersymmetric models
\cite{2}  with exact ``R--parity" predict the LSP to be stable. Notice
that supersymmetry at the weak scale has originally been introduced as
a  solution \cite{3} of the gauge hierarchy problem \cite{4} which has no
direct connection to the dark matter problem.
It is also worth mentioning that
Grand Unification of the gauge couplings of the Standard Model (SM) is only
compatible \cite{7} with recent precision measurements if
additional ``light" particles (beyond those present in the SM) exist; the
minimal supersymmetric version of the Standard Model, the MSSM \cite{2},
introduces just the right degrees of freedom to allow for Grand Unification.

In most models, the LSP is the
lightest neutralino $\chi$; they therefore automatically fulfill the tight
constraints on the cosmic density of stable charged or
colored particles that can be derived from
unsuccessful searches \cite{5} for exotic isotopes.
Moreover, model calculations
\cite{6} show that for a wide region of parameter space, the relic density
of LSPs left over from the Big Bang is just large enough to account for the
``observed" dark matter or even to allow for a flat universe as favoured by
inflationary models \cite{1}.

Nevertheless one would like to prove experimentally that some or
all of the Dark Matter is indeed made up from the lightest neutralino
$\chi$. Notice that even the discovery of some superparticle in a collider
experiment will not provide this proof; as far as collider experiments
are concerned the LSP is ``stable" if its lifetime exceeds about $10^{-8}$
seconds, while it can only contribute to DM if its lifetime is longer than
about $10^{10}$ years.

Two methods have been proposed to search for relic neutralinos (or similar
particles, like heavy neutrinos). The most direct way is to look for the
scattering of ambient DM particles off the nuclei in a detector; experiments
using Silicon and Germanium counters have already reached sufficient
sensitivity to exclude massive Dirac neutrinos or sneutrinos
as main ingredients
of the DM halo of our galaxy[8]. Alternatively one can search for energetic
neutrinos emerging from the center of the Earth or Sun. The idea here is that
a DM particle can lose energy in collisions with nuclei,
and can then be trapped
by the gravitational field of celestial bodies.
Eventually they will become
concentrated in the center of these bodies where they will annihilate.
Some fraction of these annihilation events will produce energetic (anti)
muon neutrinos that can be detected in
underground experiments. Currently the
 best bounds of this kind come from the Kamiokande group \cite{9};
in the mass range beyond 45 GeV they are more sensitive
to Dirac neutrinos than
the counter experiments, and even start to impose meaningful bounds on
Majorana neutrinos or the lightest neutralino dark matter.

A crucial ingredient in the analysis of both kinds of experiments is a good
knowledge of the elastic LSP--nucleus scattering cross section.
The counting rate in a direct detection experiment is obviously proportional
to this cross section. Since these cross sections  determine the
rate at which LSPs are captured by the Earth or Sun, they also affect the
signal rate in experiments looking for $\chi\chi$ annihilation; in the
limiting case where LSP capture by and annihilation in a given celestial body
are in equilibrium, the observable neutrino flux is again directly proportional
to the LSP--nucleus scattering cross sections.

There are two different kinds of interactions between a neutralino and a
nucleon: Those that are proportional to the spin of the nucleus, and coherent
interactions that are proportional to the nucleon number (or, approximately,
mass) of the nucleus. It is important to realize that spin dependent
interactions are not coherent, i.e. the scattering matrix element for heavy
nuclei is not enhanced compared to that of single nucleons (apart from
trivial phase space factors), and might even be suppressed by nuclear form
factors. Nevertheless the spin--dependent terms are important, since the
coherent interactions are often suppressed dynamically,
as we will see below. Indeed, early estimates
of neutralino nucleus scattering cross sections\cite{10,11} only
included the spin dependent contributions.

However, already Ref.\cite{10} pointed out that
a scalar (spin independent) LSP--quark
effective interaction does exist if the quarks are massive. This was first
studied quantitatively by Griest \cite{12},
using an effective Lagrangian approach in the limit of
negligible mixing between the superpartners of left-- and right--handed quarks;
in this case a nonzero interaction results only if the LSP is a mixture of
gaugino and higgsino states.
Srednicki and Watkins then pointed out \cite{13} that squark mixing
introduces additional terms that are generally of the same order as those
considered by Griest; these terms
survive even if the LSP is a pure gaugino or higgsino state.
Finally, Giudice and Roulet \cite{14} pointed out that Higgs exchange
also contributes to the coherent LSP--nucleus interactions,
if the LSP is not a pure state.

The effective interactions studied in Refs.\cite{12}--\cite{14}
all involve massive quarks. In case of $c$, $b$ and
$t$ quarks,  a coupling to nucleons and nuclei emerges
through heavy quark loops coupled to
two gluons \cite{15}, see Fig. 1.  However, in case of the squark exchange
contribution this entails the computation of a
loop diagram with a (squark)
propagator contracted to a point; it is a priori
not clear under which circumstances
 this treatment produces a reliable estimate.

In a previous publication \cite{16} we therefore presented a complete 1--loop
calculation of the neutralino--gluon interaction in the relevant limit of small
momentum transfer. We did indeed find contributions that are proportional
to the mass of the quark in the loop. We will demonstrate in this paper
explicitly that these terms reduce to the results of Refs.\cite{12,13} only
if the quark mass is small compared to both the squark and the LSP mass, which
is frequently not the case for the top quark. We also identified a second class
of contributions which are proportional to the LSP mass, not the quark mass;
these terms survive even in the
absence of mixing in both the neutralino and the
squark sector. This is not surprising once one realizes that the coherent
interaction is induced by the chiral symmetry breaking of the theory.
We also demonstrated that these new terms can be numerically
as important as those discussed in Refs.\cite{12,13,14}.

Two problems prevented us from fully exploiting our results in
Ref.\cite{16}. On the one hand we found that the new terms suffered from
logarithmic infrared divergences
in the limit of vanishing quark mass. On the
other hand our calculation produced terms
with gluonic operators different
from those encountered in Ref.\cite{15}. We have realized  since then that
these two seemingly distinct problems
actually have a common solution. The new
gluonic operator is nothing but a special case of the leading twist
gluonic operators encountered in analyses of deep--inelastic lepton--nucleon
scattering\cite{17}; the relevant matrix element can be deduced from
experimental results. Furthermore, the logarithmic divergence is a consequence
of mixing between leading  twist quark and gluon operators at 1--loop
level. We identified the corresponding quark operators in an expansion of the
effective LSP--quark interaction to order $\msqf$. We could then use standard
renormalization group techniques to sum the leading logarithms, which
are intimately related to ``scaling violations" observed in deep inelastic
scattering.
This enables us to write down the effective scalar (spin--independent) LSP
nucleon interaction including all terms up to order $\alpha_s\msqiv$, as
well as terms of order $\msqf$ including all leading logarithmic QCD
corrections and some non--logarithmic (``finite") corrections.

The rest of this paper is organized as follows. In Sec.2 we present the
effective LSP--quark interaction at tree level, including terms up to order
$\msqf$ for the squark exchange contribution. We identify the leading twist
quark operator that appears in this order.
In Sec.3 we discuss QCD effects. We
compare the results of Refs.\cite{12,13} with the corresponding results from
our full one loop calculation. We also carefully treat the new terms which
survive in the limit of vanishing quark mass, including the twist--2 operators
in leading logarithmic approximation as well as non--logarithmic ``trace
terms". In Sec.4 we present some numerical results for the spin--independent
LSP nucleon matrix element. We find that, as anticipated in Ref.\cite{13},
squark mixing is usually quite important; however the treatment of
Refs.\cite{12,13} fails in case of the $t$ quark contribution. We also show
some predictions of counting rate in Germanium detectors, where we include the
spin dependent contribution when it is appropriate. Sec.5 is devoted to a
brief summary of our main results. Expressions for couplings are listed in
Appendix A, while Appendix B contains some loop integrals.

\section*{2) The effective neutralino--quark interaction}

In this section we discuss the effective Lagrangian describing
neutralino--quark interactions at tree level; QCD effects will
be discussed in the next section.

Three classes of diagrams contribute to LSP--quark scattering:
The exchange of a $Z$ or Higgs boson in the $t$ channel, as well as
squark exchange in the $s$ or $t$ channel. As already mentioned in the
Introduction, LSP interactions with matter can naturally be separated
into spin--dependent and spin--independent parts. $Z$ and squark exchange
contribute to the former, and squark and Higgs exchange to the latter.
The main interest of this paper is the spin--independent interactions,
but for completeness we also list the spin dependent contribution:
\be\label{e1}
{\cal L}^{\rm{eff}}_{\rm{spin}} = \bar{\chi}\gamma^{\mu}\gamma_5
\chi \bar{q} \gamma_{\mu} (c_q +d_q \gamma_5) q, \ee
where we have defined
\ben\label{e2}\beq
c_q=&-\frac{1}{2} \sum_{i=1}^{2}\frac{a_{\tilde{q}_i} b_{\tilde{q}_i}}
{m^2_{\tilde{q}_i}-(m_{\chi} + m_q )^2}
+ \frac{g^2}{4 m_W^2} O''^R(T_{3q}-2e_q \sin^2 {\theta_W})\label{e2a};\\
d_q=&\frac{1}{4} \sum_{i=1}^{2} \frac{a_{\tilde{q}_i}^2+ b_{\tilde{q}_i}^2}
{m^2_{\tilde{q}_i}-(m_{\chi} + m_q )^2}
- \frac{g^2}{4 m_W^2} O''^RT_{3q}.
\label{e2b}\eeq\een
The second term in Eqs.(\ref{e2}) describes the $Z$ exchange contribution,
in the notation of Haber and Kane \cite{2}; $g$ is the SU(2) gauge
coupling, $\theta_W$ the weak mixing angle, $T_{3q}(=\pm 1/2)$
and $e_q$ the weak isospin and electric charge of quark $q$, and
$O''^R$ describes the $Z\chi\chi$ coupling \cite{2}.

The first term in Eqs.(\ref{e2}) is due to the exchange of the two
squarks $\tilde{q}_i$ with a given flavor, in the notation of
Ref.\cite{16}. The couplings $a_{\tilde{q}_i}$ and $b_{\tilde{q}_i}$
describe scalar and pseudoscalar LSP--quark squark interactions:
\be\label{e3}
{\cal L}_{\chi q \tilde{q}} = \sum_{i=1}^{2} \bar{q}
(a_{\tilde{q}_i} +b_{\tilde{q}_i} \gamma_5) \chi \tilde{q}_i+ h.c. \ee
It is important to realize that
$|a_{\tilde{q}_i}|=|b_{\tilde{q}_i}|$ in the chiral limit. In general these
couplings contain both gauge and Yukawa contributions, and are sensitive to
mixing in both the neutralino and squark sectors.
Neutralino mixing has been included in all recent analyses of LSP--nucleus
scattering \cite{11,12,13,14,18,19}, so we do not describe it
again. However, squark mixing has only been taken into account in
Ref.\cite{13}, so it might be useful to briefly describe it
here.

The mixing between the superpartners of
left and right handed squarks\cite{19a} is determined by the following mass
matrices \cite{19b};
\ben \label{e4} \beq
{\cal M}^2_{\tilde u} &= \mbox{$ \left( \begin{array}{cc}
m^2_{\tilde{q}_L} + m_u^2 + 0.35 D_Z & - m_u (A_u + \mu \cot \! \beta) \\
- m_u (A_u + \mu \cot \! \beta ) & m^2_{\tilde{u}_R} + m_u^2 + 0.16 D_Z
\end{array} \right) $}; \label{e4a} \\
{\cal M}^2_{\tilde d} &= \mbox{$ \left( \begin{array}{cc}
m^2_{\tilde{q}_L} + m_d^2 - 0.42 D_Z & - m_d (A_d + \mu \tanb) \\
- m_d (A_d + \mu \tanb) & m^2_{\tilde{d}_R} + m_b^2 - 0.08 D_Z
\end{array} \right) $}. \label{e4b} \eeq \een
Here, $D_Z = M_Z^2 \cos\! 2\beta$ with $\tan\!\beta $
being the usual ratio of vacuum expectation values of the two Higgs doublets,
$m_{\tilde{q}_L, \tilde{u}_R, \tilde{d}_R}$
are soft breaking masses, $A_u$ and $A_d$ are soft breaking parameters
describing the strength of trilinear scalar interactions, and $\mu$ is the
supersymmetric Higgs(ino) mass parameter that also appears in the
neutralino mass matrix. Once squark mixing is included, the mass
eigenstates $\tilde{q}_i$ become superpositions of the current
eigenstates $\tilde{q}_L, \tilde{q}_R$:
\be\label{e5}
\mbox{$\left( \begin{array}{c}
\tilde{q}_1 \\ \tilde{q}_2\end{array} \right)$}
=\mbox{$\left( \begin{array}{cc}
\cos\!\theta_q & \sin\!\theta_q\\
-\sin\!\theta_q & \cos\!\theta_q
\end{array}\right)$}
\mbox{$\left( \begin{array}{c}
\tilde{q}_L \\ \tilde{q}_R\end{array} \right)$},
\ee
with
\be\label{e6}
\sin\! 2\theta_q = \frac{2 m^2_{LR}}
{m^2_{\tilde{q}_1}-m^2_{\tilde{q}_2}},
\ee
where $m^2_{LR}$ denotes the $(1,2)$ element of the corresponding
mass matrix (\ref{e4}). In the same notation the masses of
the squark eigenstates are
\be
m^2_{\tilde{q}_{1,2}}=\frac{1}{2}
\left[m^2_{LL} + m^2_{RR}\pm \sqrt{(m^2_{LL}-m^2_{RR})^2 + 4 m^4_{LR}}
\ \right];
\ee
in our convention $m_{\tilde{q}_1} < m_{\tilde{q}_2}$.

The couplings $a_{\tilde{q}_i},b_{\tilde{q}_i}$
can now be expressed in terms of the squark mixing angles $\theta_q$
and the couplings of squark current eigenstates:
\ben\label{e7}\beq
a_{\tilde{q}_1}=\frac{1}{2}[\cos\!\theta_q (X_{q0} + Z_{q0})
+\sin\!\theta_q(Y_{q0} +Z_{q0})]
\label{e7a}; \\
b_{\tilde{q}_1}=\frac{1}{2}[\cos\!\theta_q (X_{q0} - Z_{q0})
+\sin\!\theta_q( Z_{q0} - Y_{q0})]
\label{e7b}, \eeq\een
where we have used the notation of Ref.\cite{19c}:
\ben \label{e8} \beq
X_{q0} &= - \sqrt{2} g \left[  T_{3_q} N_{02} -
\tan\!\theta_W (T_{3_q} - e_q) N_{01} \right] ; \label{e8a} \\
Y_{q0} &= \sqrt{2} g \tan\theta_W e_q N_{01}; \label{e8b} \\
Z_{u0} &= - \frac {g m_u N_{04}} { \sqrt{2} \sin \! \beta m_W}; \ \
Z_{d0} = - \frac {g m_d N_{03}} { \sqrt{2} \cos \! \beta m_W}\ . \label{e8c}
\eeq \een
In terms of the parameters $a,b,c$ of Griest \cite{12},
one has $X_{q0}= -\sqrt{2} gb$, $Y_{q0}=\sqrt{2} gc$
and $Z_{q0} = -\sqrt{2} g a$. The couplings
$a_{\tilde{q}_2}$,$b_{\tilde{q}_2}$ of the heavier
squark eigenstate can be obtained from Eqs.(\ref{e7})
by the transformation
$\sin\!\theta_q \rightarrow \cos\!\theta_q$,
$\cos\!\theta_q \rightarrow -\sin\!\theta_q$, see Eq.(\ref{e5}).

Recall that $\chi$ is a Majorana particle; the vertices one
reads off the effective Lagrangian (\ref{e1}) thus have to be multiplied with
two. Notice that we have included the LSP and quark masses in our squark
propagator; this propagator therefore accurately describes the scattering
of a massive LSP off a massive quark in the limit where the momentum transfer
is negligible.  Since ambient LSPs are expected \cite{1}
to have velocities $ v\simeq 10^{-3} c$, this approximation is justified.

Except for our refined propagator, our effective Lagrangian (\ref{e1}) agrees
with Griest's \cite{12} up to a factor 2 in the limit $\theta_q \rightarrow
0$.\footnote{We agree with his expression for the cross section.} In the same
limit we agree with Ellis and Flores \cite{18} except for the relative sign
between the $Z$ and $\tilde{q}$ exchange terms.
In the limit of small but nonzero squark mixing we reproduce the result
of Ref.\cite{13} for the squark exchange contribution, in the limit
$m^2_{\tilde{q}_i} \gg (m_q+m_{\chi})^2$.

We finally mention that the contribution from the quark vector current,
described by $c_q$, to the LSP--nucleon scattering matrix element is
suppressed by a factor of the LSP velocity $v$ and is therefore negligible in
practice; we have included it here to allow comparison with results in the
literature \cite{12,13,18}. We have however omitted squark exchange
contributions $\sim (a^2_{\tilde{q}_i}-b^2_{\tilde{q}_i})/
[m^2_{\tilde{q}_i}-(m_q+m_{\chi})^2]^2$, since they are doubly suppressed
compared to the leading terms given in Eqs.(\ref{e2}): The coupling is nonzero
only due to chirality violation, which is generally weak; moreover, this
contribution is of higher order in an expansion in inverse powers of the
squark mass.

We now turn to the spin independent neutralino--quark interactions. They
receive contributions from the exchange of squarks and scalar Higgs bosons:
\be \label{e9}
{\cal L}^{\rm{eff}}_{\rm{spin-indep.}}
=f_q\bar{\chi}\chi \bar{q}q + g_q\bar{\chi}\gamma^{\mu} \partial^{\nu}
\chi (\bar{q} \gamma_{\mu} \partial_{\nu} q -
\partial_{\nu} \bar{q} \gamma_{\mu} q).
\ee
Here we have introduced
\ben\label{e10}\beq
f_q=&-\frac{1}{4}\sum^2_{i=1}\frac
{a^2_{\tilde{q}_i}- b^2_{\tilde{q}_i}}
{m^2_{\tilde{q}_i}-(m_{\chi}+m_{q})^2}
+m_q \sum^2_{j=1}\frac{c_{\chi}^{(j)}c_{q}^{(j)}}{m^2_{H_j}};
\label{e10a}\\
g_q=&-\frac{1}{4}\frac{a^2_{\tilde{q}_i}+b^2_{\tilde{q}_i}}
{\left[
m^2_{\tilde{q}_i}-(m_{\chi}+m_{q})^2
\right]^2} .
\label{e10b}\eeq\een
The couplings $a_{\tilde{q}_i}$, $b_{\tilde{q}_i}$ entering the squark
exchange contributions are again given by Eq.(\ref{e7}). The coefficients
$c_{\chi}^{(j)}$ and $c_{q}^{(j)}$ determine the couplings of the $j$--th
scalar Higgs boson to the LSP and to the quark $q$, respectively\cite{16}; for
the convenience of the reader explicit expression are given in Appendix A.

Only the term  $\propto f_q$ has been discussed in the existing literature. In
the limit $\theta_q\rightarrow 0$ our expression (\ref{e10a}) agrees with
Ellis and Flores \cite{18} (up to an overall sign). The Higgs exchange
contribution to $f_q$ agrees with Ref.\cite{14}, where the importance of this
term was first pointed out; however, here as well as in Refs.\cite{18} and
\cite{20} only the contribution from the light Higgs boson $H_{2}$ has been
taken into account. As first pointed out in Ref.\cite{19} the contribution of
the heavier Higgs boson $H_1$ can also be very important, since its coupling
to quarks can be enhanced.\footnote{The expression for the $H_1\chi\chi$
coupling in Ref.\cite{19} contains a sign mistake; Kamionkowski, private
communication.}
However, the squark exchange contribution in Ref.\cite{19} is too large by a
factor of two. Moreover, unlike in Refs.\cite{19} and \cite{20} the relative
sign between the Higgs and squark exchange terms should not depend on the sign
of the eigenvalue $m_{\chi}$ of the neutralino mass matrix. Finally our squark
exchange contribution (\ref{e10a}) agrees with Ref.\cite{13} in the limit of
small but nonzero squark mixing.

Notice that the squark exchange contribution to Eq.(\ref{e10a}) is
proportional to the difference $a^2_{\tilde{q}_i}- b^2_{\tilde{q}_i}$, which
vanishes in the limit of a chiral LSP--quark--squark interaction. Indeed,
Eqs.(\ref{e4})--(\ref{e8}) imply that this coefficient vanishes for massless
quarks: from Eqs.(\ref{e7}) and (\ref{e8}) we see that $a^2_{\tilde{q}_i}-
b^2_{\tilde{q}_i}\propto \sin\!2\theta_q$ for massless quarks ($Z_{q0}=0$); on
the other hand, from Eq.(\ref{e6}) one has $\sin\!2\theta_q \propto
m_q/m_{\tilde{q}}$. We thus conclude that $a^2_{\tilde{q}_i}-
b^2_{\tilde{q}_i}$ is very small for the light quarks that are abundant in
nucleons. Heavy quarks couple to nucleons only through a loop; this will be
discussed in the next section. Since the $O(m_{\tilde{q}}^{-2})$ term is
suppressed, the contribution from $g_q$ can be important, even though it is
$O(m_{\tilde{q}}^{-4})$. For the interesting case of a bino--like LSP,
$a^2_{\tilde{q}_i}+ b^2_{\tilde{q}_i}$ is of the order of the squared U(1)
gauge coupling, without any suppression from small mixing angles. If the LSP
is higgsino--like, the coupling to light quarks is suppressed, but this
scenario is cosmologically not very interesting, since the relic density of
higgsino--like states is very small \cite{6,19c,21}.

In order to compute LSP--nucleus scattering cross sections
from Eq.(\ref{e9}) we have to know the nucleonic matrix elements of the
relevant quark operators. For light $(u,d,s)$ quarks the
matrix elements $\langle N|m_q \bar{q}q|N \rangle$
have to be taken from calculations
that attempt to describe strong interactions at long distances.
Following Refs.\cite{19,20} we will use results from chiral perturbation
theory as a guideline \cite{22}. The matrix elements
$\langle N|m_Q \bar{Q}Q|N \rangle$ for heavy $(c,b,t)$ quarks can be computed
as in Ref.\cite{15}, see Sec.3. Finally, the second term in
the effective Lagrangian (\ref{e9}) can be expressed in terms of twist--2 quark
operators that appear in analyses of deep--inelastic lepton--nucleon
scattering \cite{17}. This can be seen using the tensor identity
\be\label{e11}
Q_{\mu\nu}L^{\mu\nu}=(Q_{\mu\nu}-\frac{1}{4}g_{\mu\nu}Q_{\alpha}^{\alpha})
L^{\mu\nu}+\frac{1}{4}Q_{\alpha}^{\alpha} L_{\beta}^{\beta},
\ee
as well as the equation of motion for the ``trace terms"
$\bar{q}\gamma^{\mu}\partial_{\mu} q$ and $\bar{\chi}\gamma^{\nu}\partial_{\nu}
\chi$, to rewrite the second term in Eq.(\ref{e9}) as:
\be\label{e12}
{\cal L}^{\rm{eff}}_{m_{\tilde{q}}^{-4}}
=g_q \cdot \left[
-2i{\cal O}^{(2)}_{q\mu\nu}\bar{\chi}\gamma^{\mu}\partial{^\nu}\chi
-\frac{1}{2} m_q m_{\chi}\bar{q}q\bar{\chi}\chi
\right].
\ee
Here we have introduced the $n=2$ twist--2 quark operator \cite{17}
\be\label{e13}
{\cal O}^{(2)}_{q\mu\nu} =\frac{i}{2}\left[
\bar{q}\gamma_{\mu}\partial_{\nu}q+\bar{q}\gamma_{\nu}\partial_{\mu}q
-\frac{1}{2}\bar{q}\partial_{\nu}\gamma^{\nu}q
\right].
\ee
Notice that ${\cal O}^{(2)}_{q\mu\nu}$ is traceless, which is necessary for
operators with fixed spin or ``twist".

The characteristic energy scale for the effective Lagrangian (\ref{e9})
is given by the Higgs and squark propagators. In order to describe physics at
lower scales QCD renormalization effects must be included. These are the
subject of the following section.
\section*{3) QCD effects on LSP-nucleon scattering}
In this section we discuss the role QCD plays in calculating the
LSP-nucleon scattering matrix element. Two independent effects
have to be considered. On the one hand, perturbative QCD
predicts a nonzero matrix element $\langle N|m_Q \bar{Q}Q|N\rangle$
for heavy quarks
\cite{15}. On the other hand, QCD also changes the coefficient $g_q$ of the
twist-2 operator appearing in the effective Lagrangian (\ref{e9}).

We start with a discussion of the heavy quark contribution. It has
been realized more than 15 years ago that a scalar heavy quark
current couples to nucleons with strength $\sim 1/m_Q$ via the
loop diagram depicted in Fig.1. The result is \cite{15,22}:
\be\label{e14}
\langle N| m_Q \bar{Q}Q|N\rangle = \frac{2}{27} m_N(1-\sum_{u,d,s}f_{Tq}) \ \
(Q=c,b,t).
\ee
Here $f_{Tq}$ denotes the fraction of the nucleon mass
$m_N$ that is due to light quark
$q$:
\be\label{e15}
\langle N| m_q \bar{q}q|N\rangle = m_N f_{Tq} \ \ (q=u,d,s).
\ee
The numerical values of the $f_{Tq}$ have to be taken from model
calculations \cite{22}, as mentioned earlier.

Eq.(\ref{e14}) describes the Higgs exchange contribution
exactly (up to corrections of order of the momentum transfer divided by
$m_Q$); indeed, it has first been derived \cite{15} in a calculation of
the Higgs-nucleon coupling. However,
when applying Eq.(\ref{e14}) to the squark exchange
contribution, one effectively replaces a box
diagram with one squark and three quark propagators by a
triangle diagram that only contains quark propagators.
This procedure cannot be expected to yield accurate results unless
$m_{\tilde{q}}^2\gg (m_q +m_{\chi})^2$.
This is frequently not the case for stop squarks and can
also be problematic for the sbottom squark, as we will see later.

In Ref.\cite{16} we therefore presented a full 1 loop calculation of
the contribution from heavy quarks and their superpartners.
This also includes the contributions of box and triangle diagrams with two
or three squark propagators; indeed, these diagrams are necessary to
guarantee gauge invariance to all orders in $m_q/m_{\tilde{q}}$.
As mentioned above, the Higgs boson coupling to nucleons via a heavy quark
loop is given exactly by Eq.(\ref{e14}); however, there is also a contribution
involving only the superpartners of heavy quarks. The total contribution
involving squarks can be described by the following effective LSP gluon
interaction \cite{16}\footnote{Note that the signs of the terms
proportional to $T_q$, $T_{\tilde q}$, $B_{1S}$ and $B_{1D}$ were incorrect in
our original calculation.}:
\beq\label{e16}
{\cal L}^{\rm eff}_{\chi g}=&\bar{\chi}\chi F_{\mu\nu}^aF^{a\mu\nu}
\cdot \left[ -T_{\tilde{q}} +B_D +B_S \right]
\nonumber\\
&-(B_{1D}+B_{1S})\bar{\chi}\partial_{\mu}\partial_{\nu}\chi
F^{a\mu\rho}F_{\rho}^{a\nu}
\nonumber\\
&+B_{2S} \bar{\chi}(i\partial_{\mu}\gamma_{\nu}+i\partial_{\nu}\gamma_{\mu})
\chi F^{a\mu\rho}F_{\rho}^{a\nu}.
\eeq
Here $T_{\tilde{q}}$ is the Higgs contribution via squark loops, while all
other contributions come from box and triangle diagrams involving quarks
and squarks. All coefficients with a subscript D are proportional to
the difference $a^2_{\tilde{q}_i}-b^2_{\tilde{q}_i}$
(summed over quarks and squarks), while a subscript $S$ indicates a
contribution proportional to $a^2_{\tilde{q}_i} + b^2_{\tilde{q}_i}$:
\ben\label{e17}\beq
B_D=&\frac{\alpha_S}{4\pi} \frac{1}{8}\sum_{q,i}
(a^2_{\tilde{q}_i}-b^2_{\tilde{q}_i}) m_q I_1(m_{\tilde{q}},m_q,m_{\chi}),
\label{e17a}\\
B_S=&\frac{\alpha_S}{4\pi} \frac{1}{8} m_{\chi}\sum_{q,i}
(a^2_{\tilde{q}_i}+b^2_{\tilde{q}_i}) I_2(m_{\tilde{q}},m_q,m_{\chi}),
\label{e17b}\\
B_{1D}=&\frac{\alpha_S}{4\pi} \frac{1}{3}\sum_{q,i}
(a^2_{\tilde{q}_i}-b^2_{\tilde{q}_i}) m_q I_3(m_{\tilde{q}},m_q,m_{\chi}),
\label{e17c}\\
B_{1S}=&\frac{\alpha_S}{4\pi} \frac{1}{3}m_{\chi}\sum_{q,i}
(a^2_{\tilde{q}_i}+b^2_{\tilde{q}_i}) I_4(m_{\tilde{q}},m_q,m_{\chi}),
\label{e17d}\\
B_{2S}=&\frac{\alpha_S}{4\pi} \frac{1}{12}\sum_{q,i}
(a^2_{\tilde{q}_i}+b^2_{\tilde{q}_i}) I_5(m_{\tilde{q}},m_q,m_{\chi}),
\label{e17e}\\
T_{\tilde{q}}=&\frac{\alpha_S}{4\pi} \frac{1}{24}\sum_{j=1}^2
\frac{c_{\chi}^{(j)}}{m^2_{H_j}}\sum_{q,i}\frac{c^{(j)}_{\tilde{q}_i}}
{m^2_{\tilde{q}_i}}.
\label{e17f}
\eeq
\een
Expressions for the couplings appearing in Eq.(\ref{e17f}) are given in
in appendix A, while the loop integrals in Eqs.(\ref{e17a})--(\ref{e17e})
are listed in Appendix B.

The effective Lagrangian (\ref{e16}) contains terms with different tensor
structure. The matrix element (\ref{e14}) is related to the ``trace term"
$F_{\mu\nu}^a F^{a\mu\nu}$, while the terms
$\propto F^{a\mu\rho}F_{\rho}^{a\nu}$ are related to the twist--2 operator. We
will discuss these two kinds of terms in the following subsections.
\subsection*{3a) Trace Part}
We start with the contribution
$\propto F^{a}_{\mu\nu}F^{a\mu\nu}$, which corresponds to the contribution
described by Eq.(\ref{e14}). It is important to realize that we have to
include the trace part of the terms $\propto B_1$ and $ B_2$ here,
i.e. we have to apply the
tensor identity (\ref{e11}) to these terms. The total
``trace term" then becomes:
\be \label{e18}
{\cal L}^{\rm eff}_{\chi g, {\rm trace}}
=\bar{\chi}\chi F^{a}_{\mu\nu}F^{a\mu\nu}\cdot
\left[ -T_{\tilde{q}}+B_D+ B_S -\frac{m_{\chi}}{2}B_{2S}
-\frac{m_{\chi}^2}{4} (B_{1D}+B_{1S})
\right].
\ee

In order to compare Eq.(\ref{e18}) with the result one obtains from
using Eq.(\ref{e14}) on {\it all} terms $\sim m_Q \bar{Q}Q$ in the effective
Lagrangian (\ref{e9}) we use the identity \cite{15,22}
\be\label{e19}
\frac{\alpha_S}{4\pi}\langle N|F_{\mu\nu}^{a}F^{a\mu\nu}|N\rangle
= -\frac{2}{9}
m_N \left( 1-\sum_{u,d,s} f_{T_q}\right).
\ee
Moreover, we need the expansion of the loop integrals $I_{1-5}$
in powers of inverse squark mass, i.e. for the case $m_q^2\ll m_{\tilde{q}}^2
-m_{\chi}^2$:
\ben\label{e20}\beq
I_1\simeq& \frac{2}{3m_q^2(m^2_{\tilde{q}}-m^2_{\chi})}
+{\cal O} \left(\frac{1}{m^4_{\tilde{q}}}\right); \label{e20a}\\
I_2-\frac{1}{3}I_5\simeq& -\frac{2}{3m^4_{\tilde{q}}}
+{\cal O}\left( \frac{m^2_{\chi}}{m^6_{\tilde{q}}}
,\frac{m_q^2}{m^6_{\tilde{q}}}\right);\label{e20b}\\
I_4,I_5 \simeq& {\cal O}(\frac{1}{m^6_{\tilde{q}}}).\label{e20c}
\eeq\een
We see that Eqs.(\ref{e17})--(\ref{e20a}) give
the {\it same} result to leading order in $m^{-2}_{\tilde{q}}$,
as the squark exchange contribution (\ref{e10a}) to the effective Lagrangian
(\ref{e9}) when used together with Eq.(\ref{e14}). However,
Eq.(\ref{e20b}) gives a two times {\it
larger} result than Eqs.
(\ref{e10b}) and (\ref{e12}) when used together with Eq.(\ref{e14}).
This illustrates the perils of using effective Lagrangians in loop
calculations. Such a procedure can only be expected to give the
correct answer if the loop integration is dominated by momenta small
compared to the squark mass. In the limit $m_q\ll m_{\tilde{q}}$
this is true for $I_1$, which has a quadratic infrared (IR)
singularity as $m_q \rightarrow 0$.
We will see below that this is also true for $I_2$ and $I_5$
separately, which show logarithmic IR divergencies as $m_q\rightarrow 0$.
However, these singularities cancel in the relevant combination
$I_2-\frac{1}{3}I_5$.  In other words, in this particular combination  loop
momenta of the order of the squark mass make significant contributions.
The effective Lagrangian approach corresponds to introducing a
cut--off of the order of the squark mass on the loop integration; it is
clear that this cannot yield reliable results if loop momenta close to
the cut--off contribute significantly.

We thus conclude that the effective Lagrangian approach that has been used in
the literature to estimate the heavy $(c,b,t)$ quark contribution to LSP
nucleon scattering from the first term in Eq.(\ref{e9}) should give
approximately the correct result if $m_q^2\ll m^2_{\tilde{q}}-m^2_{\chi}$.
However, the heavy quark contribution from the second term in Eq.(\ref{e9})
{\it cannot} be estimated in this fashion; one has use the results of the full
1--loop calculation.

The contributions from $B_D$ and $B_S$ differ also where
light $(u,d,s)$ quarks are concerned.
As discussed earlier, their contribution from the effective Lagrangians
(\ref{e9}) and (\ref{e12}) has to be estimated using model calculations
\cite{22} for $\langle N|m_q \bar{q}q|N\rangle$.
It does not make sense to include
the light quark contribution to $B_D$, since the corresponding loop integral
$I_2$ is dominated by the nonperturbative region of small momenta;
this effect should be included in the nonperturbative nucleonic matrix
elements. On the other hand, the light quark contribution to $B_S
-\frac{m_{\chi}}{2}B_{2S}$ {\it can } safely be included, since here the
relevant loop integral is dominated by the perturbative region of large
momenta, which is not included in the matrix elements of $m_q \bar{q}q$.
Another way to see that these two contributions are indeed independent
and can thus safely be added is the observation that in Eq.(\ref{e12})
chirality is violated by the quark mass, while in $B_S- \frac{m_{\chi}}{2}
B_{2S}$ the LSP mass provides the source of chirality breaking.

\subsection*{3b) Contributions from twist--2 operators}
We now turn to a discussion of QCD effects on the twist--2 quark operator
(\ref{e13}) in the effective Lagrangian (\ref{e12}). As mentioned at the end of
section 2, the effective Lagrangian describes physics at the energy
scale given by the squark propagator. In more formal language
 we have to know the matrix element
$\langle N|{\cal O}^{(2)}_{q\,\mu\nu}|N\rangle$ at the renormalization point
$\mu_0 =\sqrt{m^2_{\tilde{q}}-m_{\chi}^2}$ in order to
directly use Eq.(\ref{e12}). This matrix element is closely
related to the second moment of quark distribution functions\cite{17}:
\be\label{e21}
\langle N|{\cal O}^{(2)}_{q\mu\nu}|N\rangle \left|_{\mu_0} \right. =
\frac{1}{m_N}(p_{\mu}p_{\nu}-\frac{1}{4}m_N^2 g_{\mu\nu})
\int^1_0 dx \, x \cdot \left[
q(x,\mu_0^2) +\bar{q}(x,\mu^2_0)
\right],
\ee
where $p$ is the nucleon momentum\footnote{
Recall that we assume negligible momentum transfer, so that the initial and
final momentum are identical.}.

In principle we could use Eq.(\ref{e21}) directly. However this has the
practical disadvantage that the hadronic matrix elements would depend on SUSY
parameters, via the scale dependence of the parton distribution functions
(these are the famous scaling violations of QCD). We therefore choose to
express our result in terms of matrix elements of twist--2 operators at the
fixed (low) scale $Q_0=5$ GeV ($\simeq m_b$). The moments of parton
densities at high momentum scales are related to those at a lower scale
via renormalization group equations \cite{17}:
\ben\label{e22}\beq
\frac{d}{d\ln Q}q(n,Q^2)&=-\left[
\gamma_{FF}^F(n) q(n,Q^2)+\gamma_{VV}^{F}(n) G(n,Q^2)
\right];\label{e22a}\\
\frac{d}{d\ln Q}G(n,Q^2)&=-\left[
\gamma_{FF}^V(n) \sum_q q(n,Q^2)+\gamma_{VV}^{V}(n) G(n,Q^2)
\right],\label{e22b}
\eeq\een
where we have introduced
\be\label{e23}
q(n,Q^2)\equiv \int^1_0 dx x^{n-1}q(x,Q^2),
\ee
and similarly for $G(n,Q^2)$. Notice that the sum in
Eq.(\ref{e22b}) runs over all quarks and antiquarks whose mass is
(much) smaller than $Q$. The $\gamma_{jj}^{i}(n)$ are components of a $2
\times 2$ matrix of anomalous dimensions \cite{17}.

In order to solve Eqs.(\ref{e22}) we first observe that flavor nonsinglet
operators $q_i(n,Q^2)-q_j(n,Q^2)$, where $i$, $j$ are flavor indices,
renormalize multiplicatively, since the gluonic contribution in
Eq.(\ref{e22a}) cancels out. It is always possible to express $l$
independent quark densities in terms of $l-1$ differences and the total sum
$\Sigma (n,Q^2)\equiv \sum_{\rm quarks} q(n,Q^2)$. Finally, in order to
treat the mixing between $\Sigma (n)$ and $G(n)$ one introduces orthogonal
combinations ${\cal O}_+ (n)$ and ${\cal O}_{-}(n)$  which again renormalize
multiplicatively with anomalous dimensions $\gamma_+(n)$ and $\gamma_{-}(n)$.
For the relevant case $n=2$ and $N_f=5$ light flavors of quarks,
these operators are given by \cite{17}
\ben\label{e24}\beq
{\cal O}_+(2,Q^2) = &\frac{16}{31}\Sigma(2,Q^2)-\frac{15}{31}G(2,Q^2),
\label{e24a}\\
{\cal O}_-(2,Q^2) = &\frac{15}{31}\left[\Sigma(2,Q^2)+G(2,Q^2)\right],
\label{e24b}
\eeq\een
while the relevant anomalous dimensions are
\be\label{e25}
\gamma_{FF}^F(2)=\frac{16\alpha_s}{9\pi};\ \
\gamma_{+}(2)=\frac{31\alpha_s}{9\pi};\ \ \gamma_{-}(2)=0.\ee
Notice that the second moment of a parton density is nothing
but the fraction of the total nucleon momentum carried by that species of
partons. The sum over all
quarks and gluons $\Sigma(2) + G(2)$ must therefore be equal to 1 at all
momentum scales.
This explains why $\gamma_-(2)$ vanishes. We thus succeeded in reducing the
system of coupled equations (\ref{e22}) to decoupled equations of the form
\be\label{e26}
\frac{d}{d\ln Q}{\cal O}(Q^2) = -\frac{\alpha_S}{\pi}
\tilde{\gamma} {\cal O}(Q^2),
\ee
which has the solution \cite{17} (for $N_f=5$ flavors)
\be\label{e27}
{\cal O}(Q^2)= {\cal O}(Q_0^2)
\left( \frac{\alpha_S(Q^2)}{\alpha_S(Q_0^2)}
\right)^{6\tilde{\gamma}/23}.
\ee
Here we have used the standard expression for the running QCD coupling
constant:
\be\label{e27a}
\alpha_S(Q^2) =\frac{12\pi}{23 \ln \frac{Q^2}{\Lambda^2}} .
\ee

In order to apply Eqs.(\ref{e22})--(\ref{e27a}) to our problem, we first
observe
that we
can safely ignore the Yukawa contributions to the combinations of couplings
$a^2_{\tilde{q}_i}+b^2_{\tilde{q}_i}$ for $u,d,s$ and c (s)quarks.
If we further assume that squarks of the first two generations are
degenerate in mass, as suggested by the analysis of SUSY contributions to
meson mixing in the $K^0$ and $B^0$ systems \cite{23} , we find that the
coefficients $g_q$ are equal for $u$ and $c$ quarks, and for $d$ and $s$
quarks. On the other hand, for large $\tan\beta$ the bottom Yukawa coupling
can be sizable; moreover, if $\tan\beta$ and/or $\mu$ are large
Eq.(\ref{e4b}) leads to non--negligible mass splitting between sbottom squarks.
In general $g_b$ can therefore differ significantly from $g_d = g_s$.
Altogether we therefore need 3 independent combinations of
quark densities for 5 quarks (The contribution of the top quark will be
discussed later):
\beq\label{e28}
\sum_{\rm quarks}g_q q(Q^2)=&
\frac{1}{5}(2g_u+2g_d+g_b)\left[{\cal O}_+ (Q^2)+{\cal O}_- (Q^2)\right]
\nonumber\\
&+\frac{1}{2}(g_u-g_d)\left[ {\cal O}_u(Q^2)-{\cal O}_d(Q^2) \right]
\nonumber\\
&+\frac{1}{5}(g_u+g_d-2g_b)\left[ \frac{1}{2}\Sigma (Q^2)-5 b(Q^2)\right].
\eeq
Here we have used Eqs.(\ref{e24}) to replace the sum over quark
densities $\Sigma$ by the orthogonal combinations ${\cal O}_+$ and
${\cal O}_-$.
Moreover we have made the usual assumption that $q(x)=\bar{q}(x)$
for the $s,c$ and $b$ quark densities in nucleons, and have introduced
\be\label{e29}
{\cal O}_u=u+\bar{u}+2c; \ \ {\cal O}_d=d+\bar{d}+2s.
\ee

Eq.(\ref{e28}) holds at any scale $Q^2$. As discussed earlier, we have to
evaluate it at the high scale $\mu_0^2=m_{\tilde{q}}^2-m_{\chi}^2$,
but we want to express the result in terms of parton densities at fixed scale
$Q_0^2=25$ GeV$^2\simeq m_b^2$; this scale has been chosen such that
$b(Q_0^2)=0$.\footnote{Nevertheless the last term in
Eq.(\ref{e28}) renormalizes
multiplicatively, like any other nonsinglet quark density, since $b(Q^2)$
is nonzero for $Q^2>Q_0^2$.} Using Eqs.(\ref{e25})--(\ref{e27}) we have
\beq\label{e30}
\sum_{\rm quarks}g_q q(\mu_0^2)=
&\frac{1}{5}(2g_u+2g_d+g_b)\cdot
\left[\frac{15}{31}+{\cal O}_+(Q_0^2)
\left( \frac{\alpha_s(\mu_0^2)}{\alpha_s(Q^2_0)}\right)^{62/69} \right]
\nonumber\\
&+\left\{ \frac{1}{2}(g_u-g_d) \left[ {\cal O}_u(Q^2_0)
-{\cal O}_d(Q^2_0) \right]\right. \nonumber\\
&\ \ \ \ \ \left.+ \frac{1}{10}(g_u +g_d -2g_b) \Sigma(Q^2_0)\right\}
\left( \frac{\alpha_s(\mu_0^2)}{\alpha_s(Q^2_0)} \right)^{32/69},
\eeq
where we have used momentum conservation, which implies ${\cal O}_-(2)=15/31$
as discussed above. The other necessary combinations of moments of parton
densities at scale $Q_0^2$ can easily be obtained by integrating standard
parametrizations of these densities. Results for four recent representative
parametrizations \cite{24,25} are collected in table 1.

Eq.(\ref{e30}) describes our results in terms of constant
parameters $g_q$ of the effective Lagrangian and scale--dependent hadronic
matrix elements. Equivalently we could have used running couplings and constant
matrix elements. The crucial observation here is that the product
$g_q(Q^2)\cdot q(Q^2)$ is {\it independent} of the scale $Q^2$.

Notice that Eq.(\ref{e30}) depends on the gluon density at scale $Q_0^2$.
This dependence is entirely due to the scale dependence of the parton
densities,
which in turn is induced by QCD loops. If we expand Eq.(\ref{e30})
in $\alpha_s$, keeping only terms linear in $\alpha_s$, the coefficient of
$G(Q_0^2)$ on the right hand side of Eq.(\ref{e30})
is $(\alpha_S /
3\pi) \ln(\mu_0/Q_0) \cdot \sum g_q$. On the other hand, the traceless
part of the combination of gluon field strength in the second and third
term in the effective Lagrangian (\ref{e16}) is nothing but the $n=2$
twist--2 gluon operator:
\be\label{e31}
\langle N|F^{a\mu\rho}F^{a\nu}_{\rho}
+\frac{1}{4} g^{\mu\nu}F^{a}_{\alpha\beta}
F^{a\alpha\beta}|N\rangle =
\frac{1}{m_N}(p^{\mu}p^{\nu}-m_N^2 g^{\mu\nu} ) G(2).
\ee
Moreover, in the limit $m_q\rightarrow 0$ the coefficient
$B_{2S}$ contains a logarithmic divergence:
\beq\label{e32}
B_{2S}\simeq &\frac{\alpha_S}{4\pi}\frac{1}{6}
\sum_{q,i}\frac{ a^2_{\tilde{q}_i} + b^2_{\tilde{q}_i} }
{(m^2_{\tilde{q}_i} - m_{\chi}^2)^2 }
\ln \frac{m^2_{\tilde{q}_i} - m^2_{\chi}}{m_q^2} +{\rm finite\ terms}
\nonumber\\
=&-\frac{\alpha_S}{3\pi}\ln \frac{\mu_0}{m_q}\sum g_q;
\eeq
in the second step we have assumed squarks to be (essentially)
degenerate, as for the discussion leading to Eq.(\ref{e30}).
The renormalization group analysis thus exactly reproduces the
leading logarithmic contribution to the full
1--loop amplitude\footnote{ Recall that the twist--2 quark operators appear
with coefficient --2 in the effective Lagrangian (\ref{e12}), while the term
$\propto B_{2S}$ in Eq.(\ref{e16}) gets a factor of 2 since we have used the
explicitly symmetric form of the LSP tensor, following the notation of
Ref.\cite{16}.} if we identify $Q_0$ with the quark mass $m_q$. This
is no surprise; in the given case the loop integral is again
dominated by small loop momenta, leading to an IR divergence as
$m_q\rightarrow 0$, so we expect the effective Lagrangian approach to
(approximately) reproduce the 1--loop result.

For the case of light ($u,d,s$) quarks we cannot trust the 1--loop
calculation, since their masses are so small that perturbative QCD is no
longer trustworthy at scales $Q \simeq m_q$. Non--perturbative effects become
important, which are included in the (measured) parton distribution
functions. The light quarks therefore have to be treated using the effective
Lagrangian (\ref{e12}) and (\ref{e30}).

On the other hand, for heavy $(c,b,t)$ quarks the perturbative result
of Eqs.(\ref{e16}) and (\ref{e17}) should be reliable to the given (1--loop)
order. Nevertheless the use of the RGE approach, Eq.(\ref{e30}),
might be
advantageous, since it automatically sums leading logarithmic QCD corrections
to all orders\cite{17}, i.e. all terms $\sim (\alpha_s \ln\frac{\mu_0}{m_q})
^n$ are included. For the case of $c$--quarks,
we find that the logarithmic term
does indeed dominate the loop integral $I_5$. Summing higher powers of this
logarithm is therefore quite important; we thus
decided to treat $c$--quarks in
the RGE approach. In case of the top quark the
logarithmic term in $I_5$ dominates only if the stop squark is very heavy,
in which case the total contribution $\sim g_t$ is very small anyway.
It therefore appears more important to include non--logarithmic
terms, rather than re--summing higher powers of the logarithm. We
thus treat the $t$--quark contribution in exact 1--loop order, using Eqs.
(\ref{e16}), (\ref{e17}) and (\ref{e31}), where we take the gluon density at
scale\footnote{
At high momentum scale $G(n,Q^2)$ depends only very weakly on $Q^2$. The
use of a fixed $Q^2$ is therefore justified for top quark mass
in the allowed range between 100 and 200 GeV.}
$Q^2= 2\cdot 10^4$ GeV$^2 \simeq m_t^2$.

The case of the (s)bottom contribution is somewhat more ambiguous. For
small $\tan\beta$, $\tilde{b}$ squarks are usually
degenerate with
the other squarks, and $I_5$ is dominated by the logarithmic term. However,
for
large $\tan\beta$ the lighter sbottom eigenstate can be much lighter than the
other squarks, and can even be close in mass to the LSP. In this case the
exact 1--loop treatment seems more appropriate. In particular $I_5$ remains
finite as $m_{\tilde{b}_1}\rightarrow m_{\chi}$ or
$m_{\tilde{b}_1}\rightarrow m_{\chi}+m_b$, while the coefficient $g_b$
diverges in this limit. We therefore decided to treat the (small)
contribution from the heavier sbottom $ \tilde{b}_2$ together with the squarks
of the first 2 generations via Eq.(\ref{e30}).
In order to include the contribution from $\tilde{b}_1$ exchange we introduce
a new scale $\mu_b^2 = m^2_{\tilde{b}_1}-m^2_{\chi}$; as discussed above, this
scale can be significantly smaller than $\mu^2_0$.
We then compare the $\tilde{b}_1$ exchange contribution
according to Eqs.(\ref{e16}) and (\ref{e17})
with the prediction from the RGE approach,
and adopt the {\it smaller} of the two results as
our best estimate. Since the terms of higher order in $\ln\frac{\mu_b}{Q_0}$
tend to reduce the 1--loop result, this procedure implies that we treat the
$\tilde{b}_1$ contribution via the RGE method if $\tilde{b}_1$ is heavy.
However,
close to the spurious pole in $g_b$ the RGE prediction diverges, and our
procedure automatically chooses the 1--loop result.

This completes our treatment of QCD effects. We now turn to a discussion of
the numerical importance of the various contributions to LSP--nucleon
scattering.
\section*{4) Numerical results}
\subsection*{4a) The LSP--nucleon scattering amplitude}
Before we can give numerical results for LSP--nucleon scattering
amplitudes we have to specify the values of some parameters.
As already mentioned in the previous section, recent predictions
\cite{24,25} for the second moments of the relevant combination of
parton densities are listed in table 1.  We see that existing data
on deep--inelastic lepton--nucleon scattering, and related data which
enter the fits of Refs.\cite{24,25}, are sufficient to pin down these
parameters to better than 10\% accuracy.

We also need to specify the values of
$f_{Tq}\equiv \langle N|m_q\bar{q}q|N\rangle /m_{N}$
for light $(u,d,s)$ quarks. The contributions from $u$ and $d$ quarks is
quite small, but we include them for completeness. The combination
$\frac{m_u +m_d}{2}\langle N|\bar{u} u + \bar{d}d|N\rangle$ is
determined from the $\pi N$ ``sigma term".  However, since $u$ and $d$
quarks couple differently, we have to know the individual contributions, which
are more model dependent. Following Cheng\cite{22}, we take
\beq\label{e33}
&f_{Tu}^{(p)}=0.023;\ \ \  f_{Td}^{(p)}=0.034;\nonumber\\
&f^{(n)}_{Tu}=0.019;\ \ \  f_{Td}^{(n)}=0.041,
\eeq
where the superscripts denote protons $p$ and neutrons $n$.

The strange quark contribution is expected to be larger than that
from $u$ and $d$ quarks but the exact value is quite uncertain.
Values as small as 71 MeV \cite{26} and as high as 430 MeV\cite{22}
have been given for the matrix element $\langle N|m_s\bar{s}s|N\rangle$.
Here we follow the
recent analysis of Gasser et al.\cite{27}:
\be\label{e34}
f_{Ts}=0.14;
\ee
it should be kept in mind that this value is uncertain to
about a factor of 2, however.

In order to estimate the contribution from the spin--dependent
interactions of Eqs.(\ref{e1}) and (\ref{e2}) we have to know the matrix
elements
\be\label{e35}
\langle N|\bar{q}\gamma_{\mu}\gamma_5 q |N\rangle =2 s_{\mu} \Delta q.
\ee
Here $s_{\mu}$ is the spin vector of the nucleon, and $\Delta q$
denotes the second moment of the polarized quark density \cite{28}.
Just like the unpolarized parton densities, the $\Delta q$ can be extracted
from analyses of deep--inelastic lepton nucleon scattering. However in this
case

both probe and target have to be polarized, which complicates the experiments
significantly. Analyses of old SLAC data \cite{29} and of more recent
data from the EMC collaboration \cite{30} suggest \cite{28}:
\be\label{e36}
\Delta u =0.77,\ \  \Delta d = -0.49,\ \  \Delta s =-0.15
\ee
for the polarized quark densities in the proton; in case of the neutron
the $u$ and $d$ quark densities have to be interchanged as usual.
The errors of the quantities (\ref{e36}) are about $\pm 0.08$.
Very recently, data on polarized lepton--neutron scattering
have become available\cite{31}. A recent analysis \cite{32} of these
data gives values for the $\Delta q $ very close to those in Eq.(\ref{e36}),
but with somewhat reduced errors.

We are now in a position to present numerical results for LSP--nucleon
scattering. We begin with a discussion of the squark exchange contributions to
the spin--independent interaction via the $f_q$ term in Eq.(\ref{e10})
as well as via the $B_D$ terms in Eq.(\ref{e16}).
As mentioned in the introduction, this was the first spin--independent
contribution to be studied quantitatively\cite{12,13}, using effective
Lagrangian techniques. We write the scalar contribution to the effective
LSP--nucleon interaction as
\be\label{e37}
{\cal L}^{\chi N}_{{\rm scalar}} = f \bar{\chi}\chi \bar{\Psi}_N \Psi_N,
\ee
where $\Psi_N$ denotes the nucleon N. In the treatment of
Ref.\cite{12,13} the squark exchange contribution to the coefficient $f$
is
\be\label{e38}
f_D^{\rm eff}= m_N\left[\sum_{u,d,s} \frac{f_q^{(\tilde{q})}}
{m_q}f_{Tq}+ \frac{2}{27}f_{TG}\sum_{c,b,t}\frac{ f_q^{(\tilde{q})}}{m_q}
\right]
\ee
where $f_{TG}=1-\sum_{u,d,s}f_{Tq}$. The coefficients $f_q^{(\tilde{q})}$
are the squark exchange contributions to Eq.(\ref{e10a})
\footnote{Recall that $f_q/m_q$ is finite as $m_q\rightarrow 0.$ },
and the subscript $D$ indicates that we are only including terms proportional
to the difference of couplings $a^2_{\tilde{q}}-b^2_{\tilde{q}}$ here.
In contrast, we treat the contribution from heavy quarks in exact 1--loop
approximation as described in Sec.3:
\be\label{e39}
f_{D}= m_N\left[
\sum_{u,d,s}\frac{f_q^{(\tilde{q})}}{m_q}f_{Tq}
-\frac{8\pi}{9\alpha_S}f_{TG}(B_D-\frac{m_{\chi}^2}{4}B_{1D})\right].
\ee

Predictions from Eq.(\ref{e38}) and (\ref{e39}) are compared in Fig.2.
Here and in the following figures we have chosen a common soft breaking mass
$m_{\tilde{q}}$ for all squarks, but we include squark mass splitting due
to ``D--terms" as well as L--R squark mixing, see Eqs.(\ref{e4}). We
also assume the usual unification relation \cite{2}
$M_1 =5/3\tan^2 \theta_W M_2$ between the SUSY breaking U(1) and SU(2)
gaugino masses. Finally, in this ``global SUSY" scenario we assume that
$m_{\tilde{q}}$, $M_2$, the Higgs(ino) mass parameter $\mu$, the ratio
of vevs $\tan\beta$, the trilinear soft breaking parameter $A$
(which we also assume to be the same for all squarks), the pseudoscalar Higgs
mass $m_P$ and the top mass $m_t$ can all be varied independently.
In Fig.2 we have chosen $m_{\tilde{q}}=M_2=200$ GeV, $m_P= $500 GeV,
$m_t=140$ GeV, $A=0$ and $\tan\beta=10$.

The solid curves show the predictions of our ``exact" treatment
(\ref{e39}); the upper curves show the total contribution, while the lower
curves show the top (s)quark contribution alone. The long
dashed curve have been obtained from Eq.(\ref{e38}) including squark mixing in
the approximate treatment of
Ref.\cite{13}, while the long--short dashed curves correspond to the treatment
of Ref.\cite{12} where squark mixing was ignored. Finally, the dotted
curves show the total spin--independent contribution, as discussed in more
detail below.

We observe that the effect of squark mixing is quite important
everywhere, except at small values of $|\mu|$ where $\chi$ is higgsino--like
and the expected relic density $\Omega h^2$ very small. Their importance of
squark mixing grows with $|\mu|$ since the off--diagonal entries of the squark
mass matrices (\ref{e4}) are $\propto \mu$. On the other hand, the
approximate treatment of Ref.\cite{13} can be brought into good agreement
with our ``exact" result for the total $f_D$.
To achieve such a good agreement we have modified the squark propagators in
$f_q$, Eq.(\ref{e10a}), slightly, setting $m_q=0$ for all quarks; moreover,
in order to avoid a spurious pole in the top contribution, we have replaced the
stop propagators simply by $1/m^2_{\tilde{t}}$. This treatment is rather
ad hoc, of course, but neither omitting $m_{\chi}^2$ in all squark propagators
(as in the original treatment of Ref.\cite{13}), nor including it everywhere
gives nearly as good an approximation to the full result.
The lower curves show that this treatment still overestimates the (s)top
contribution significantly, since the loop integral $I_1$ is strongly
suppressed for $m_q> m_{\chi}$; this effect cannot be treated properly in the
framework of Ref.\cite{12} and \cite{13}. Nevertheless it affects the total
$f_D$ only in the comparatively
uninteresting case of a higgsino--like LSP.

Finally, Fig.2 shows that the inclusion of squark mixing reduces
dependence of $f_D$ on $\mu$. From Eqs.(\ref{e7}), (\ref{e8}) and (\ref{e10a})
we see that $f^{(\tilde{q})}_q$ gets contributions from higgsino--gaugino
mixing
($\propto X_{q0} Z_{q0}, Y_{q0} Z_{q0}$), as well as from squark mixing
$(\propto \sin 2\theta_q X_{q0} Y_{q0})$. The former contribution decreases
with increasing $|\mu|$, while the latter increases. Altogether one can write
schematically for a bino--like LSP and small squark mixing angle $\theta_q$:
\be\label{e40}
f_q^{(\tilde{q})}\sim \frac{g^2 \tan\theta_W}
{2(m^2_{\tilde{q}}-m_{\chi}^2)} \left[
c_1 \frac{y_L+y_R}{2} \frac{m_q}{\mu+M_2} +
c_2 \frac{m^2_{LR}}{m^2_{LL}+ m^2_{RR}} y_L y_R \tan\theta_W
\right],
\ee
where $m_{LL}^2$ etc. are again elements of the squark mass
matrices (\ref{e4}), $y_{L,R}$ are hypercharges
 and $c_1$ and $c_2$ are numbers of order 1.

In Fig.3 we show various Higgs exchange contributions to the parameter $f$
of Eq.(\ref{e37}):
\beq\label{e41}
f_H=\left(\sum_{u,d,s}\frac{f_q^{(H)}}{m_q}f_{Tq}+
\frac{2}{27}f_{TG}\sum_{c,b,t}\frac{f_q^{(H)}}{m_q}\right) m_N
+\frac{8\pi}{9\alpha_S}f_{TG}m_N T_{\tilde{q}}.
\eeq
Here, $f_q^{(H)}$ is the Higgs exchange contribution to the
coefficient $f_q$ of Eq.(\ref{e10a}), and $T_{\tilde{q}}$ appears in the
effective Lagrangian (\ref{e16}). In Fig.3 we have fixed
$m_{\tilde{q}}=-\mu=300$ GeV, $m_P= 500$ GeV, $m_t=140$ GeV and $A=0$
and varied $\tan\beta$. The leading radiative corrections from the top--stop
sector to the masses and
mixing angle of the scalar Higgs bosons \cite{33} have been taken into account.

For the given choice of parameters squarks are quite heavy, and
hence the contribution from squark loops (short dashed--dotted line)
is very small; we found previously \cite{16} that it can become sizable only
for nonvanishing $A$. In Fig.3 the heavy scalar Higgs boson $H_1$ is
approximately degenerate with the pseudoscalar, and hence is even heavier than
the squarks. Nevertheless $H_1$ exchange dominates the total Higgs
contribution for $\tan\beta\geq 5$. There are two reasons for this:
Firstly, the $H_2\chi\chi$ coupling (A2b)
goes through zero at $\tan\beta\simeq 6$,
due to a cancellation between two terms. Secondly, in the relevant limit
$m_P^2 \gg m_Z^2$, the coupling of $H_2$ to quarks are almost identically to
that of
the SM Higgs boson, independent of $\tan\beta$. On the other hand, the
couplings of $H_1$ to down--type quarks are enhanced at large $\tan\beta$.
Finally we note that for our choice (\ref{e34}) for the strange quark matrix
element $f_{Ts}$ the light quark contribution is actually larger than the one
from heavy quark loops.
This is especially true for $H_1$ exchange, since here effectively only
down--type quarks contribute once $\tan\beta>3$ or so. We have to keep
in mind, however, that the contribution from light quarks is uncertain to a
factor of 2 or so, while the heavy quark contribution depends only weakly on
$f_{Ts}$.

As already emphasized in the Introduction, Higgs bosons can only
couple to the LSP if it has both higgsino and gaugino components. These
couplings will therefore be suppressed for large $|\mu|$ and/or $|M_2|$,
since in this limit the LSP becomes an almost pure state \cite{6,19c}.
The overall order of magnitude of the Higgs exchange contribution for a
bino--like LSP can be estimated as
\be\label{e42}
f_q^{(H)}\sim \frac{g^2 \tan^2{\theta_W}}{4} \frac{m_q}{M_2+\mu}
\left(
\frac{c_3}{m^2_{H_2}}+\frac{c_4(\tan\beta)^{-2I_{3q}}}{m^2_{H_1}}
\right),
\ee
where $c_3$ and $c_4$ are again numerical factors of order 1, and the
exponent in the second term describes the enhancement (suppression)
of the $H_1$ coupling to up(down) quarks.

In Fig.4 we show contributions to the $\chi N$ interaction parameter $f$
of Eq.(\ref{e37}) that are proportional to the sum of couplings
$a^2_{\tilde{q}_i}+b^2_{\tilde{q}_i}$; none of these contributions have been
taken into account in the existing literature. In order to write
these contributions in the form of Eq.(\ref{e37})
we use Eqs.(\ref{e21}) and (\ref{e31}) as well as the identities
\ben\label{e43}\beq
\bar{\chi}\gamma_{\mu}\partial_{\nu}\chi&= -\frac{i}{m_{\chi}}k_{\mu}
k_{\nu}\bar{\chi}\chi; \label{e43a}
\\
\left( p_{\mu}p_{\nu}-\frac{1}{4}m_N^2 g_{\mu\mu}\right) k_{\mu}k_{\nu}
&=\frac{3}{4}m^2_Nm_{\chi}^2,\label{e43b}
\eeq\een
where $p$ and $k$ are the momenta of the nucleus $N$ and LSP $\chi$
respectively, and we have omitted terms proportional to the LSP velocity
$v$. The total result is:
\beq\label{e44}
f_S=\frac{1}{2} m_{\chi}m_N \sum_{u,d,s}g_q f_{Tq}
+\frac{8\pi}{9\alpha_S}\left( B_S-\frac{m_{\chi}}{2} B_{2S}
-\frac{m^2_{\chi}}{4} B_{1S}
\right) f_{TG}m_N
\nonumber\\
+\frac{3}{2}\left[ B_{2S}^{t}+\frac{m_{\chi}}{2}(B_{2S}^{t}+ B_{2S}^{t})
m_N m_{\chi} \right] G(m_t^2)
-\frac{3}{2}\sum_{u,d,s,c,b}g_q q(\mu_0^2) m_N m_{\chi}.
\eeq
The first term comes from the trace over the
twist--2 quark operators, see Eq.(\ref{e12}).
The second term originates from the trace of the twist--2 gluon
operator, as discussed in Sec.3a.  The third term is the (s)top (s)quark
contribution to the gluonic twist--2 operator; as discussed in Sec.3b
we alway treat this contribution in exact 1--loop approximation. The last term
in Eq.(\ref{e44}) gives the light quark contribution to twist--2 operators,
evaluated according to Eq.(\ref{e30}). Recall that for small mass difference
$m_{\tilde{b}_1}-m_{\chi}$ the contribution from the lighter sbottom eigenstate
is also treated in exact 1--loop order, as discussed in detail at the end of
Sec.3b. In Fig.4 we therefore show the $\tilde{b}_1$ contribution
together with $\tilde{t}$ contribution.
We see from Fig.4 that the last term in Eq.(\ref{e44}) dominates,
followed by the second term. The other two contribution are usually
negligible. The overall order of magnitude of $f_S$ for a bino--like LSP is
\be\label{e45}
f_S\sim\frac{g^2 \tan^2\theta_W}{4}\sum_q \frac{y_q^2}{(m_{\tilde{q}}^2-
m_{\chi}^2)^2 }q(2)\cdot m_{\chi}m_N.
\ee
Comparison with Eq.(\ref{e40}) shows that $f_S$ falls off faster
with increasing squark mass than the contribution $\propto
 a^2_{\tilde{q}_i} -b^2_{\tilde{q}_i}$. This is also demonstrated by the
short dashed--dotted curve in Fig.4, which shows the total contribution
from squark exchange. At small  $m_{\tilde{q}}$ the contributions are
compatible; moreover, since here $f_q^{(\tilde{q})}$ is dominated by
the second term in Eq.(\ref{e40}), the curves also have similar slopes.
However for heavier squarks the first term dominates Eq.(\ref{e40}),
and the total squark exchange contribution drops less rapidly with
increasing  $m_{\tilde{q}}$ than $f_S$ does.
\subsection*{4b) Counting rate in a Germanium detector}
Having discussed the relative importance of various contributions to the spin
independent LSP--nucleon scattering amplitude, we are now in a position to
present numerical results for elastic LSP scattering off heavy target nuclei.
We choose Germanium isotopes for this discussion, since a new class of
Germanium detectors with increased sensitivity for DM searches is expected to
become operational in the near future \cite{34}. The relevant quantity for
direct search experiments is the interaction rate which is usually measured in
events/(kg$\cdot$day). It is given by
\be\label{e46}
R= \frac{\sigma\cdot\xi}{m_{\chi}m_A}\cdot \frac{1.8\cdot 10^{11} {\rm GeV}^4}
({\rm kg\cdot day)}\cdot\frac{\rho_{\chi}}{0.3 ({\rm GeV/cm^3})}
\cdot\frac{\bar{u}_{\chi}}{320 ({\rm km/sec})}.
\ee
Here, $m_A$ is the mass of the nucleus under consideration, $\rho_{\chi}$
is the {\it local} LSP mass density, and $\bar{u}_{\chi}$
the average or effective LSP density that results from integrating over the
(assumed) Maxwellian velocity distribution of the LSP \cite{35}.
The elastic LSP nucleon cross section for the idealized case of a pointlike
nucleon is given by
\be\label{e47}
\sigma= \frac{4m_{\chi}^2m_A^2}{\pi(m_{\chi}+m_A)^2}
\left\{
\left[ Z f_p + (A-Z)f_n \right]^2
+4 \lambda^2 J(J+1) \left[\sum_{u,d,s} d_q \Delta q\right]^2
\right\}. \ee
Here $Z$ and $A$ are the charge and isotope number of the nucleus;
recall that the contribution $\propto f_{Tu}, f_{Td}$ as well the
$u$-- and $d$--quark densities are different for protons and neutrons. $J$
is the total spin of the nucleus, and $\lambda$ is a nucleonic matrix element
that basically describes the fraction of the total spin that is due to the
spin (rather than the orbital angular momentum) of the nucleons. As
usual in the literature\cite{10,11,12,18,19} we will assume that only
a single ``valence"nucleon contributes to the total spin, so that
the sum in the second term of Eq.(\ref{e47}) runs over the polarized parton
densities of either a proton or a neutron. We refer the interested reader to
Ref.(\cite{36}) for a more comprehensive treatment of the nuclear physics of
LSP--nucleus scattering. For our case of $^{73}$Ge, we adopt the value
\cite{18}$\lambda^2 J(J+1)=0.065$.

Finally, the factor $\xi$ in Eq.(\ref{e46}) describes the suppression due to
nuclear form factors. Ellis and Flores \cite{18} found that Gaussian form
factors describe the total counting rate adequately. For the standard
assumptions about the LSP velocity distributions one has \cite{18}
\be\label{e48}
\xi=\frac{0.573}{B}\left\{ 1-\frac{\exp[-B/(1+B)]}{\sqrt{1+B}}
\frac{{\rm erf}\left(\sqrt{1/(1+B)}\right)}{{\rm erf}(1)}
\right\}
\ee
where erf is the error function, and
\be\label{e49}
B\equiv \frac{m_{\chi}^2 m_A^2}{(m_{\chi}+m_A)^2}\cdot
\frac{8}{9} r^2 \bar{v}_{\chi}^2,
\ee
with $\bar{v}_{\chi}\simeq \bar{u}_{\chi}/1.2$ being the ``velocity
dispersion" \cite{35}.
For the spin--independent interaction we use
$r=r_{{\rm charge}}=(0.3+0.89A^{1/3})$
fm, while for the spin--dependent interaction of $^{73}$Ge we use a
slightly softer form factor \cite{18} with $r_{\rm spin}=1.25 r_{{\rm
charge}}$.\footnote{Our expression of counting rate (\ref{e46}), (\ref{e47})
is four times larger than the result given in Ref.\cite{18}. As stated
earlier, the scattering amplitudes that one derives from the effective
Lagrangians (\ref{e1}) and (\ref{e37}) have to include a factor of 2, due to
the Majorana nature of the LSP. Our spin--dependent cross section (\ref{e47})
agrees with Ref.\cite{12}. }

In Fig.5a--c we show contours of constant counting rate in the
plane $(M_2, m_{\tilde{q}})$, where $M_2$ is the SU(2) gaugino mass
and $m_{\tilde q}$ the SUSY breaking contribution to squark masses, which
we assume to be identical for all squarks. We have fixed
$\mu=m_P=300$ GeV and take a top mass of 140 GeV. In Fig.5a,b
we have chosen $\tan\beta=2$, while Fig.5c is for $\tan\beta=8$.
In all figures the central region between the dotted lines is excluded by
constraints from unsuccessful sparticle searches at LEP\cite{37}, and
the regions below the dotted lines at large $|M_2|$ and small
$m_{\tilde{q}}$ are excluded by the requirement $m_{\chi}\leq m_{\tilde{t}_1}$.
Finally, in the hatched regions relic LSPs would overclose the universe
($\Omega h^2>1$), while in the shaded region the LSP relic abundance is too
small to account for a significant fraction of the observed dark matter
($\Omega h^2<0.05$).

In Figs.5 a,b we observe large differences
between the regions of positive and
negative $M_2$. (More precisely only the sign of the product
$\mu\cdot M_2 \cdot\tan\beta$ is relevant here.) For given $|M_2|$, the
LSP is slightly lighter and has a considerably larger higgsino component if
$M_2>0$. This enhances the couplings to Higgs and $Z$ bosons, and
increases the first contribution to $f_q^{(\tilde{q})}$ in Eq.(\ref{e40}),
which is due to neutralino mixing. Furthermore, there is a
cancellation in the $H_2\chi\chi$ coupling for $M_2<0$ \cite{19c}.
As a result the expected counting rate is generally larger and
the LSP relic density smaller for positive $M_2$. Unfortunately this
correlation implies that often combinations of parameters that lead to a
large counting rate for fixed local LSP density also lead
to such a low universal relic abundance that the LSP does not make a good DM
candidate any more. This is demonstrated by the region of large positive
$M_2$ in Figs.5 a,b, where almost the entire region with counting rate
$>0.1$ event/(kg$\cdot$day) also has relic density $\Omega h^2< 0.05$.
This correlation has previously been observed in Ref.\cite{38}, where
only the light Higgs boson exchange contribution was taken into account.
On the other hand, for the given choice of parameters there are
large regions in the $(M_2,\mu)$ plane where the LSP does make an interesting
DM candidate and the counting rate is larger
than 0.02 event/(kg$\cdot$day),
if $M_2>0$; for $M_2<0$ the counting rate is at most a few times $10^{-3}$,
events/(kg$\cdot$day) unless $m_{\tilde{q}}$ is close to $m_{\chi}$.

Comparison of Figs. 5a (for $^{76}$Ge) and 5b (for $^{73}$Ge)
shows that the spin--dependent contribution to the cross section is
quite small. In most cases the $^{73}$Ge counting rate is
even slightly smaller than the $^{76}$Ge rate, due to the smaller coherence
enhancement factor. The only exception occurs in the region of small
$m_{\tilde{q}}$ and small negative $M_2$, where the LSP is photino--like
and the spin--dependent squark exchange contribution is sizable.
In this particular case the $^{73}$Ge counting rate can be as much as 50
\% larger than the $^{76}$Ge rate. However, usually the
difference in counting rate is of order 10\% or less.
A pure--isotope detector does therefore not seem to offer much
of an advantage over a detector using the natural mix of Ge isotopes.

At $\tan\beta=8$, Fig. 5c, the region with large counting
rates has shrunk considerably.  Even in the half--plane of
positive $M_2$ the counting rate now often falls below
0.02 events/(kg$\cdot$ day). One reason is
that increasing $\tan\beta$ from 2 to 8 increases the mass of the
light Higgs boson by approximately 50\%. In addition, bino--higgsino mixing is
suppressed (enhanced) compared to the case $\tan\beta=2$
for positive (negative) $M_2$. Both effects
conspire to reduce the
counting rate and increase the relic density
for $M_2>0$. In addition, for
$M_2<0$ the contributions from the exchange
of the heavy and light Higgs
scalar often tend to cancel, see Fig.3;
this explains why the region with
very small counting rate, below $2\cdot 10^{-3}$, is much larger
in Fig.5c than in Fig.5a. On the other hand, large $\tan\beta$
also enhances $\tilde{b}$ and $\tilde{s}$ squark mixing and reduce
$\tilde{c}$ squark mixing, see Eqs.(\ref{e4}). Since the
$\tilde{b}$ and $\tilde{c}$ contributions tend to cancel for
$M_2<0$, this results in an increase of $f_D$ in Eq.(\ref{e39}),
by more than a factor of 5. This explains the
growth of the region with counting rate$\geq 0.02$
events/(kg$\cdot$ day) in the half--plane with negative $M_2$.
Finally we remark that at large $|\tan\beta|$ the sign of $M_2
\cdot\mu\cdot\tan\beta$ is less important than for small
$|\tan\beta|$; in the limit $|\tan\beta|\rightarrow \infty$,
this sign becomes irrelevant.

So far we have only considered ``global SUSY" models where
$m_{\tilde{q}}$, $m_P$, $M_2$, $\mu$, $\tan\beta$, $A$ and $m_t$
can all be varied independently. From the theoretical
point of view ``minimal Supergravity", or ``SUGRA" models
\cite{2} are more attractive.
These models do not only allow to describe potentially realistic
sparticle spectra with fewer parameters, they can also explain
(rather than parametrize) electroweak gauge symmetry breaking in
terms of radiative corrections to the Higgs potential.
In these models supersymmetry breaking is described by
a common scalar mass $m_0$, a common gaugino mass $m_{1/2}$,
and a common $A$ parameter. This degeneracy is assumed to be
exact only at some very high energy scale, e.g. the unification scale
$M_X$. At lower scales the masses of different scalars and different
gauginos differ due to quantum corrections,
which can be described by a set of coupled renormalization
group equations (RGE) \cite{39}. In the remaining two figures
we have used simple analytical parametrizations \cite{19b,19c}
of the exact numerical solutions of these equations.
Moreover, we have required the correct amount of gauge
symmetry breaking, i.e, the correct $W$ mass at
\setcounter{footnote}{0}
the weak scale. This leaves us with \
altogether 5 free parameters\footnote{Unlike in Ref.\cite{19b}
we do not assume $B(M_X)=A-m_0$
here.}, which we take to be the SU(2) gaugino mass
$M_2$ at the weak scale, $m_t$, $A$, $m_0$
and $\tan\beta$. Notice that
$\mu$ and $m_P$ are derived quantities in this scheme.

In Fig.6 we show contours of constant counting rate in a
$^{76}$Ge detector in the $(M_2, m_0)$ plane, for $m_t=140$ GeV,
$A=0$ and $\tan\beta$=2. We immediately see that the
counting rate is much smaller than in Figs.5. One reason
is that now squarks of the first two generations are
at least five times heavier than the LSP. This is because the
RGE lead to a positive contribution $\simeq 8 M_2^2$ to
squared on--shell squark masses, while we still have
$|m_{\chi}|\leq |M_1|\simeq |M_2|/2$. As a result, the total
squark exchange contribution is usually just a few \% or even less
of the Higgs exchange contribution. Therefore Fig.6 shows
no regions of large counting rate for large $|M_2|$ and
small $m_0$, although some combinations
of parameters are still excluded by
demanding the LSP to be electrically neutral; the light charged
sparticle in this case is the light $\tilde{\tau}$ state, however,
which is much lighter than the lightest squark \cite{19b}.

Another reason for the small counting rate is that we now always have
$\mu\geq |M_2|$, since $\mu$ is fixed by the
condition to get correct symmetry breaking.
Therefore
$\chi$ is always an almost pure gaugino state in this
figure, usually a bino. The lines of constant
counting rate almost coincide with lines of constant $M_2$;
large $|M_2|$ implies larger $\mu$ and smaller
$\chi\chi H$ couplings, see Eq.(\ref{e42}). For the given
choice of $m_t$, $A$ and $\tan\beta$, $\mu$ increases
very slowly with $m_0$. Larger values of $m_0$ also
imply larger values of the light Higgs boson mass,
because of increasing top--stop loop corrections \cite{33}.
These two effects explain the $m_0$
dependence of the counting rate. Finally, the
mass of the peudoscalar Higgs boson also increases with $|M_2|$
and $m_0$. As a result $m_P$ is always well above $2m_{\chi}$,
and there is no region of small relic density at large $|M_2|$.
Instead, we find an unacceptably large relic density at
large $m_0$ and $|M_2|\geq 100$ GeV, because
the dominant slepton exchange
diagrams are suppressed here; see Ref.\cite{19c}
for a discussion of the
LSP relic density in minimal SUGRA.

The results of Fig.6 do not depend strongly on $m_t$
and $A$. Larger $m_t$ implies larger $\mu$ and less neutralino mixing
as well as larger Higgs boson masses, leading to smaller counting
rates, but for $M_2<0$ the rate only changes by a factor of 2
or so when $m_t$ is increased to 170 GeV or reduced to 110 GeV.
For $M_2>0$ the dependence on $m_t$ is stronger, since
the higgsino--component of $\chi$ is not only larger here,
it also increases faster with decreasing $\mu$. Similar
remarks also apply for the $A$--dependence. If $A\cdot M_2>0$,
both $\mu$ and $m_{H_{2}}$ increase with $|A|$, and the
counting rate decreases. However, very large values of $|A|$
are excluded because the lighter $\tilde{t}$ squark
would become too light.

On the other hand the counting rate does depend quite strongly
on $\tan\beta$, as illustrated in Fig.7. Since the cross
section is always dominated by Higgs exchange contributions, the
overall shape of the curves resembles
the square of the solid curve in Fig.3. However, there
are some important differences. In the region of small
$\tan\beta$, $\mu$ increases rapidly with $\tan\beta$ \cite{19b},
which reduces the $H\chi\chi$ couplings.
This partly compensates the reduction
of $m_{H_{2}}$ at small $\tan\beta$. Once $\tan\beta\geq 5$
or so, $\mu$ becomes almost independent of $\tan\beta$. However,
$m_P$ and hence the mass of the heavier Higgs boson $H_1$
decrease at large values of $\tan\beta$; recall that
$H_1$ exchange is much more important than $H_2$ exchange
once $\tan\beta>15$.\footnote{This also implies that
the total counting rate in this region of parameter space strongly
depends on the value $f_{Ts}$ of the hadronic matrix element
$\langle N| m_s \bar{s}s|N\rangle$, which is uncertain to a factor of 2 or so.}
Near the end of the curves squark exchange contributes about 10--15
\% of the total scattering amplitude, due to enhanced Yukawa couplings
and enhanced squark mixing in the $s$, $b$ quark sector.
Together with the reduction of $m_{H_{1}}$
this implies a faster increase of the counting  rate than what
one would expect from Fig.3, where $m_P$ was taken constant.
Eventually $m_P$ is reduced to a value close to $2m_{\chi}$,
and the LSP relic density drops below 0.05\cite{19c};
we have terminated the curves at this point.
Finally, we see that for large $\tan\beta$ the counting
rate does depend on $m_0$. This is because $m_{H_{1}}$
increases with $m_0$. However this dependence is
clearly still much weaker than
the dependence on $\tan\beta$.
\section*{5) Summary and conclusions}
In this paper we presented a detailed discussion of
elastic LSP--nucleon scattering, with emphasis on the role played by
strong interactions. We have little to add to the present
understanding of the spin--dependent contribution to the scattering
amplitude, because the operators $\bar{q}\gamma_{\mu}\gamma_5
q$ that appear in the spin--dependent effective LSP--quark interaction of
Eq.(\ref{e1}) are not renormalized by strong interactions if $m_q=0$.
In other words the quantities $\Delta q$ that one introduces to
parametrize the hadronic matrix elements of these operators do not depend on
the renormalization scale.

On the other hand, QCD effects are crucial for the understanding of
spin--independent LSP--nucleon interactions. Such interactions would be absent
in a world where chirality is conserved exactly. In the case at hand,
chirality breaking can enter either via the quark mass or via the LSP mass. In
previous studies \cite{12,13,14,18,15} only terms $\propto m_q$ were included.
Both Higgs and squark exchange contribute to these terms; the hadronic matrix
elements $\langle N|m_Q \bar{Q}Q|N\rangle$ that relate LSP--quark scattering
to LSP--nucleon scattering were treated using the result of Ref.\cite{15}. We
argued in Sec.3a that this is strictly speaking not correct for the squark
exchange contribution, since it involves the evaluation of a loop integral
with one propagator contracted to a point. We were nevertheless able to find a
modification of this ``effective" treatment of Refs.\cite{12,13} that usually
reproduces the full 1--loop calculation\cite{16} for the total heavy quark
contribution quite accurately. However, this treatment fails for the top quark
contribution. For the cases we checked this changed the total squark exchange
contribution by more than $\sim$20\% only if the total amplitude was dominated
by the Higgs exchange terms. However, in models where the (s)top contribution
is enhanced this ``effective" treatment may no longer be sufficient. Finally we
emphasize that mixing between the superpartners of left-- and right--handed
quarks is usually important \cite{13} whenever the squark exchange
contribution is sizable. Squark mixing is as generic  a prediction of SUSY
models as LSP mixing is; ignoring it can therefore give quite misleading
results.

The existence of terms where chirality is broken by the LSP mass rather than
the quark mass has previously been noted by us \cite{16}. In Sec.2 and Sec.3b
of this paper we discussed the connection between these terms and the
"twist--2" operators that appear in analyses of deep inelastic lepton--nucleon
scattering. We used the QCD--improved parton model which automatically
re--sums leading logarithmic corrections to all order in perturbation theory.
This is quite important, reducing the contribution from $b$ and $c$ (s)quarks
to these terms by a factor of two or more, compared to the 1--loop result. We
emphasize again that this class of contributions survives even in the limit of
no squark mixing and no mixing in the neutralino sector, unlike the other
coherent contributions. On the other hand, this new contribution is
proportional to $m_{\tilde{q}}^{-4}$, while the previously discussed squark
exchange contribution is $m_{\tilde{q}}^{-2}$. Within the MSSM this
contribution is therefore only numerically important if squarks are not much
heavier than the LSP.

It should be kept in mind that all squark exchange contributions are usually
subdominant if squarks are very heavy. If we simply multiply all mass
parameters in the neutralino and squark sectors of the theory with a constant
factor (keeping $m_W$ fixed), all coherent squark exchange contributions scale
with the inverse third power of this factor; this can most easily be seen from
our approximate expressions (\ref{e40}) and (\ref{e45}) given in sec.4a. We
also find that the spin--dependent amplitude decreases as the inverse second
power of this overall mass scale. The Higgs sector behaves differently,
because it has to give the correct symmetry breaking. The heavy Higgs exchange
contribution again falls with the third power of this scale. However, the
light Higgs boson exchange contribution (\ref{e42}) only decreases linearly,
due to reduced neutralino mixing. This comes from the well--known fact
\cite{33} that the mass of the light Higgs boson is bounded from above in the
MSSM, almost independently of the overall sparticle mass scale. Similar bounds
can be derived even in non--minimal models \cite{41}. We thus find that for
sufficiently heavy sparticles, light Higgs exchange will always dominate the
total LSP--nucleus scattering cross section. Of course increasing all
sparticle masses arbitrarily not only leads to fine--tuning problems, it can
also easily lead to an unacceptably large cosmological relic density \cite{6}.

In Sec.4b we used our LSP--nucleon scattering amplitude to compute relic
neutralino scattering rates in a Germanium detector. We found that these rates
depend quite strongly on the parameters of the model, as indicated by the
above scaling law. Even the relative sign of parameters of the neutralino mass
matrix was found to be very important. The proposed detectors \cite{34} aim
for a sensitivity of about 0.1 event/(kg$\cdot$day). We see from Figs.5 that
such a large counting rate can usually only be expected if either the light
Higgs boson or the LSP itself is quite light; both cases should be testable at
the second phase of the LEP collider. In order to arrive at this conclusion we
exclude combinations of parameters that lead to a very small overall LSP relic
density, since in such a situation the LSP is obviously not a good DM
candidate. We also find that usually, although not always, the spin--dependent
interaction is too small to be detectable.

The situation simplifies somewhat in the more restrictive minimal supergravity
models. Unfortunately here the expected counting rate is usually quite low.
Indeed, we often are in the "asymptotic" region discussed above, where only
the light Higgs exchange contribution survives, unless the ratio of vacuum
expectation value $\tan\beta$ is large, in which case the exchange of the
heavier Higgs boson makes the dominant contribution. The least favorable
situation occurs at intermediate values of $\tan\beta$ where the counting rate
could fall well below $10^{-3}$ event/(kg$\cdot$day) even for an only
moderately heavy sparticle spectrum, as seen in Fig.7 where $m_{\chi}\simeq$
100 GeV.

Given that most combinations of parameters give a counting rate well below 0.1
event/(kg$\cdot$day), should we pursue this kind of experiment further? We
think the answer to this question is emphatically `yes'. First of all, we do
not know how Nature choose her parameters. There are regions in parameter
space not excluded by any experiment that do lead to sizable counting rates.
Moreover, as already mentioned in the Introduction, a positive relic LSP
signal would yield information that {\it cannot} be obtained from any collider
experiment. For instance, a positive signal would immediately give a lower
bound of the order of $10^{10}$ years on the LSP lifetime, some 25 orders of
magnitude beyond what can be achieved at collider experiments. Moreover, such
a signal would obviously greatly enhance our knowledge of how our galaxy has
formed.

On the other hand, even if these experiments did reach a sensitivity of
$10^{-3}$ or even $10^{-4}$ events/(kg$\cdot$day) they would not preempt the
motivation for SUSY searches at colliders. This is because the absence of a
signal in an LSP search experiment is difficult to translate into stringent
bounds on model parameters. We already mentioned that an unstable but
long--lived LSP could never be detected in this fashion; this could easily be
accommodated by introducing R--parity breaking interactions at a strength well
below possible experimental limits. Furthermore, one has to realize that the
expression (\ref{e46}) for the counting rate depends on several parameters
beyond those describing the sparticle spectrum. We already mentioned the
sizable uncertainty in the strange quark matrix element $\langle N|m_s
\bar{s}s|N\rangle$, which can result in as much as a factor of 4 uncertainty
in the counting rate. Perhaps even more worrisome is the uncertainty in the
local neutralino flux, $\rho_{\chi}\cdot\bar{u}_{\chi}$ in Eq.(\ref{e46}).
Existing studies \cite{42} conclude that this is known to a factor of 2 or so,
based on current models of galaxy formation. We are no experts in this field,
but to our knowledge no ``standard model of galaxy formation" has emerged yet.
It should be noted here that the currently accepted best guess value for the
local LSP density $\rho_{\chi}$ is some 5 orders of magnitude larger than the
universal relic density. Similar, current estimates of the velocity
$\bar{u}_{\chi}$ are some 7 orders of magnitude above the thermal velocity of
Big Bang relics. Together, this gives a ``galactic enhancement factor"
(compared to the average over the Universe) of $\sim 10^{12}$. Clearly LSP
detection would be hopeless without this enhancement, but can present galaxy
formation models really predict this factor up to a factor of 2 or so?

Our conclusion is therefore that sparticle searches at colliders are
complementary to LSP detection experiments. Each kind of experiment can yield
information not accessible to the other. The results presented in this paper
should allow for the as yet most accurate calculation of the cross sections
relevant for the analysis of experiments searching for cosmic relic
neutralinos.

\subsection*{Acknowledgements}
We thank K. Hikasa, M. Savage, K. Griest and M. Luke for very useful
discussions and suggestions, and X. Tata and D. Zeppenfeld for discussions.
M.D. thanks the particle physics group at the University of Hawaii and
especially X. Tata for their hospitality during his visit where part of this
work was completed; after having walked the streets of Waikiki at night, he
begins to understand why someone with this (inverted) first name likes Hawaii.
 This work was supported in part by the U.S. Department of Energy under
contract No. DE-AC02-76ER00881, and in part by the Wisconsin Research
Committee with funds granted by the Wisconsin Alumni Research Foundation. The
work of M.D. was supported by a grant from the Deutsche Forschungsgemeinschaft
under the Heisenberg program.

\renewcommand{\theequation}{A.\arabic{equation}}
\setcounter{equation}{0}

\section*{Appendix A}
In this Appendix we give explicit expressions for the
$Z$ and Higgs couplings that enter the
effective Lagrangians of Eqs.(\ref{e1}) and (\ref{e9}).
We use the notation of Ref.\cite{19c},
which is very similar to the conventions of Haber and Kane \cite{2} and Gunion
and Haber \cite{43}. In particular, we denote the four components of the
eigenvector of the $4 \times 4$ neutralino mass matrix that corresponds to the
LSP (the lightest neutralino) by $N_{0i}$; $i=1$ corresponds to the bino
component, $i=2$ the $SU(2)$ gaugino component, and $i=3$ and $i=4$ are the
hypercharge $Y=-1/2$ and $+1/2$ higgsino components, respectively.
Note that we take $N_{0i}$ to be real; this means that the sign of
the eigenvalue $m_{\chi}$ has to be kept when evaluating the scattering
amplitudes. However the kinematical prefactor in the expression (\ref{e47})
for the cross section only depends on the kinematical mass, i.e. the absolute
value of $m_{\chi}$.
Mixing in
the Higgs sector is described \cite{43} by the angles $\beta$ and $\alpha$;
\tanb\ is the ratio of the vevs of the $Y=+1/2$ and $Y=-1/2$ neutral Higgs
bosons, and $\alpha$ describes the mixing of the neutral scalar mass
eigenstates.

We are now in a position to list the couplings appearing in the main text. We
start with the
LSP--$Z$ coupling in Eqs.(\ref{e2}):
\be\label{a1}
O"^R=\frac{1}{2}(N_{04}^2-N_{03}^2).
\ee
Expressions for the LSP-quark-squark couplings have already been given in the
main text, Eqs.(\ref{e7}),(\ref{e8}).

We next turn to the LSP--Higgs couplings appearing in Eqs.(\ref{e10a})
and (\ref{e17f}). They are given by \cite{43}:
\ben \label{a2} \beq
c^{(1)}_{\chi} &= \frac{1}{2} ( g N_{02} - g' N_{01} ) ( N_{04} \sin \! \alpha
- N_{03} \cos \! \alpha); \label{a2a} \\
c^{(2)}_{\chi} &= \frac{1}{2} ( g N_{02} - g' N_{01} ) ( N_{03} \sin \! \alpha
+ N_{04} \cos \! \alpha), \label{a2b} \eeq \een
where $g$ and $g'$ are the $SU(2)$ and $U(1)_Y$ gauge coupling, respectively.
Recall that the superscript 1 refers to the heavier scalar Higgs boson.

The Higgs--quark couplings entering Eq.(\ref{e10a}) can be
written as \cite{43}
$c^{(i)}_q = g\cdot r^{(i)}_q/(2 m_W)$, with:
\be \label{a3}
r^{(1)}_u = - \frac {\sin \! \alpha}{\sin \! \beta}; \ \
r^{(1)}_d = - \frac {\cos \! \alpha}{\cos \! \beta}; \ \
r^{(2)}_u = - \frac {\cos \! \alpha}{\sin \! \beta}; \ \
r^{(2)}_d = \frac {\sin \! \alpha}{\cos \! \beta}; \ee
here $u \ (d)$ stands for any charge = $+2/3 \ (-1/3)$ quark.

Finally the diagonal Higgs--squark couplings appearing
in Eq.(\ref{e17f}) can be written as
(remember that $\tilde{q}_1 = cos  \theta_q \tilde{q}_L + \sin \! \theta_q
\tilde{q}_R$ stands for the lighter squark mass eigenstate):
\ben \label{a4} \beq
c^{(i)}_{\tilde{q}_1} &= \frac{g m_Z} {\cos \! \theta_W} s^{(i)}
( I_{3_q} \cos^2 \theta_q - e_q \sin^2 \theta_W \cos \! 2 \theta_q )
+ \frac {g m_q^2} {m_W} r^{(i)}_q \nonumber \\
& - \frac {g m_q \sin \! 2 \theta_q} {2 m_W} (A_q r_q^{(i)} + \mu
r'^{(i)}_q); \label{a4a} \\
c^{(i)}_{\tilde{q}_2} &= \frac{g m_Z} {\cos \! \theta_W} s^{(i)}
( I_{3_q} \sin^2 \theta_q + e_q \sin^2 \theta_W \cos \! 2 \theta_q )
+ \frac {g m_q^2} {m_W} r^{(i)}_q \nonumber \\
&+ \frac {g m_q \sin \! 2 \theta_q} {2 m_W} (A_q r_q^{(i)} + \mu
r'^{(i)}_q). \label{a4b} \eeq \een
The parameters $A_q$ and $\mu$ enter the off--diagonal elements of the
squark mass matrices of Eq.(\ref{e4}), $m_W$ and $m_Z$ are the masses of the
wea
k
gauge bosons, $\theta_W$ is the weak mixing angle, $I_{3_q} = \pm 1/2$ and
$e_q$ are the third component of the weak isospin and electric charge of quark
$q$, respectively, and the $r_q^{(i)}$ have already been defined in
Eq.(\ref{a3}); finally, the $s^{(i)}$ and $r'^{(i)}_q$ are given by:
\ben \label{a5} \beq
s^{(1)} &= - \cos (\alpha + \beta); \ \
s^{(2)} = \sin (\alpha + \beta); \label{a5a} \\
r'^{(1)}_u &= - \frac {\cos \! \alpha} {\sin \! \beta}; \ \
r'^{(1)}_d = - \frac {\sin \! \alpha} {\cos \! \beta}; \ \
r'^{(2)}_u =   \frac {\sin \! \alpha} {\sin \! \beta}; \ \
r'^{(2)}_d = - \frac {\cos \! \alpha} {\cos \! \beta}. \label{a5b} \eeq \een

\renewcommand{\theequation}{B.\arabic{equation}}
\setcounter{equation}{0}

\section*{Appendix B: Loop integrals}
In this appendix we list the loop integrals that
appear in Eqs.(\ref{e17a})--(\ref{e17e}):
\ben \label{b1} \beq
I_1(\msq,m_q,\mc) &= \int_0^1 dx \frac {x^2 - 2x + 2/3} {D^2} \label{b1a}\\
&= \frac {1} {\Delta} \left[ \frac {\mqsq-\mcsq}{3 \msqsq} - \frac {2}{3}
\frac {\msqsq - \mcsq}{\mqsq} - \frac{5}{3} + \left(2 \msqsq - \frac{2}{3}
\mcsq \right) L \right]; \nonumber \\
I_2(\msq,m_q,\mc) &=  \int_0^1 dx \frac {x(x^2 - 2x + 2/3)} {D^2}
\nonumber \\
&= \frac {1}{2 m^4_{\chi}} \left[ \ln \frac {\msqsq} {\mqsq} - \left(
\msqsq -\mqsq - \mcsq \right) L \right] \nonumber \label{b1b} \\
&+ \frac {1} {\Delta} \left\{ \left[ \frac {m_q^4 - \mqsq \msqsq} {\mcsq}
- \frac {7}{3} \mqsq + \frac{2}{3} (\mcsq - \msqsq) \right] L \right.
\nonumber \\ & \left. \hspace*{1cm} + \frac {\mqsq -\mcsq} {3\msqsq}
+ \frac {\msqsq - \mqsq} {\mcsq} + \frac {2}{3} \right\}.
\\
I_3(\msq,m_q,\mc) &=  \int_0^1 dx \frac {x^2(1-x)^2} {D^3}
\nonumber\\
&=\frac{3(\mcsq-\mqsq-\msqsq)}{\Delta^2}+ \frac{L}{\Delta}
\left( -1 + \frac{6\mqsq\msqsq}{\Delta}\right) ;\label{b1c}
\\
I_4(\msq,m_q,\mc) &=  \int_0^1 dx \frac {x^3(1-x)^2} {D^3}
\nonumber \label{b1d}\\
&=\frac{1}{2 m_{\chi}^6}\left[ \ln\frac{\msqsq}{\mqsq}-
(\msqsq-\mqsq-m^4_{\chi})L \right] - \frac{1}{\msqsq\mcsq}
\\
&-\frac{\mqsq(\msqsq-\mqsq-\mcsq)}{m_{\chi}^4 \Delta}L
+\frac{1}{\Delta}\left[ \frac{\mqsq}{m_{\chi}^4}
-\frac{1}{\msqsq}\left( 1-\frac{\mqsq}{\mcsq} \right)^2 + \frac{1}{2\mcsq}
\right]
\nonumber\\
&+\frac{3\mqsq}{\Delta^2}
\left\{ 1+\frac{\msqsq-\mqsq}{\mcsq}
+ \left[ \frac{\mqsq(\mqsq-\msqsq)}{\mcsq}-2\mqsq-\msqsq -\mcsq \right] L
\right\} ; \nonumber \\
I_5(\msq,m_q,\mc) &=  \int_0^1 dx \frac {x(1-x)(2-x)} {D^2}
\nonumber\\
&=\frac{1}{2 m_{\chi}^4}
\left[ \ln \frac{\msqsq}{\mqsq} -(\msqsq-\mcsq-\mqsq)L \right]
\\
&-\frac{1}{\Delta}\left\{ L\cdot \left[
2(\msqsq-\mcsq)+3\mqsq+\frac{\mqsq(\msqsq-\mqsq)}{\mcsq}
\right]
-3+\frac{\mqsq-\msqsq}{\mcsq}\right\}.\label{b1e} \nonumber
\eeq \een
Here we have introduced the quantities
\ben \label{b2} \beq
D &= x^2 \mcsq + x \left( \msqsq- \mqsq - \mcsq \right) + \mqsq;
\label{b2a} \\
\Delta &= 2 \mcsq \left (\mqsq + \msqsq \right) - m^4_{\chi} -
\left( \msqsq - \mqsq \right)^2; \label{b2b} \\
L &= \frac {2} {\sqrt{|\Delta|}} \arctan \frac {\rt} {\mqsq + \msqsq - \mcsq},
\ \ \ \ \ \ \Delta \geq 0, \nonumber \\
 &= \frac {1} {\sqrt{|\Delta|}} \ln \frac {\mqsq + \msqsq - \mcsq + \rt}
{\mqsq + \msqsq - \mcsq - \rt} \ \ \
\Delta \leq 0.
\label{b2c} \eeq \een
Eqs.(\ref{b1}) are only valid if $m_{\tilde{q}}>m_{\chi}$, which is always
true in our case. Notice also that $I_{1-5}$ are finite as either
$m_{\chi}\rightarrow 0$ or $\Delta\rightarrow 0$. The first case can be
treated numerically in a straightforward fashion. The limit
$\Delta\rightarrow 0$
can be treated by expanding Eqs.(\ref{b1}),(\ref{b2}) around
the point $m_{\chi}=|m_{\tilde{q}}-m_q|$. Alternatively,
this limit can be treated numerically by setting $\Delta$ to some
(small) constant $\delta$ as $\Delta\rightarrow 0$, and
taking $m^2_{\chi} =\mqsq+\msqsq-\sqrt{4\mqsq \msqsq -\delta}$.

\clearpage
{\bf Table:} Numerical values of the second moments of the combinations of
parton densities appearing in eq.(\ref{e38}). ``MTLO'', ``MTB1'' and ``MTB2''
refers to the leading order parametrization and two next--to--leading order
parametrizations of ref.\cite{24}, where we have set $b(Q_0^2) \equiv 0$
exactly, while ``Owens'' refers to the parametrization of ref.\cite{25}.
Columns 2 to 5 refer to scale $Q_0^2 \simeq m_b^2$, while column 6 gives the
gluon density at scale $Q^2 = 2 \cdot 10^4 \ {\rm GeV}^2 \simeq m_t^2.$ The
superscripts in columns 3 and 4 refer to the proton and neutron; the other
entries are identical for both kinds of nucleon.
\begin{center}
\begin{tabular}{|c||c|c|c|c|c|}
\hline
  & ${\cal O}_+$ & ${\cal O}_u^p - {\cal O}_d^p$ &
${\cal O}_u^n - {\cal O}_d^n$ &  $ \Sigma $ & $G$ \\
\hline
MTB1 & 0.025 & 0.115 & --0.181 & 0.509 & 0.492 \\
MTB2 & 0.056 & 0.102 & --0.190 & 0.540 & 0.471 \\
MTLO & 0.019 & 0.113 & --0.181 & 0.503 & 0.480 \\
Owens & 0.018 & 0.109 & --0.171 & 0.502 & 0.514\\
\hline
\end{tabular}
\end{center}
\newpage
\section*{Figure Captions}

\renewcommand{\labelenumi}{Fig.\arabic{enumi}}
\begin{enumerate}

\item 
On the left we depict the Feynman diagrams that contribute to
spin--independent LSP--quark interactions. In the low--energy limit and after
a Fierz rearrangement of the $\tilde{q}$ exchange contribution these give rise
to the effective interaction depicted in the center. In previous work the
contribution involving $c,b,t$ quarks has been estimated closing the quark line
as shown in the diagram at the right.

\vspace*{5mm}

\item 
Squark exchange contributions $\propto a^2_{\tilde{q}_i}-b^2_{\tilde{q}_i}$ to
the spin--independent effective coupling $f$ of Eq.(\ref{e37}). The solid and
long--dashed lines show results from our ``exact" treatment (\ref{e39}) and
the traditional approach (\ref{e38}), respectively; the upper (lower) set of
curves is for the total (top--stop) contribution. The long--short dashed lines
show the prediction from Eq.(\ref{e38}) if squark mixing is neglected. The
dotted curves depict the total result for $f$, including terms $\propto
a^2_{\tilde{q}_i}+b^2_{\tilde{q}_i}$  and Higgs exchange contributions. The
curves are not extended into the experimentally excluded region of $|\mu|$.

\vspace*{5mm}

\item 
Higgs exchange contributions to the effective coupling $f$ of Eq.(\ref{e37}),
see Eq.(\ref{e41}). The solid curve shows the total Higgs exchange
contribution, the long dashed and long--short dashed curves show the light
Higgs exchange contributions involving light($u,d,s$) and heavy ($c,b,t$)
quark, respectively, and the short dashed and dot--long dashed curves depict
the corresponding heavy Higgs exchange contributions. The dot--short dashed
curve shows the contribution from the last term in Eq.(\ref{e41}), while the
dotted curve again depicts the total result for $f$.

\item 
Squark exchange contributions $\propto a^2_{\tilde{q}_i}+b^2_{\tilde{q}_i}$ to
the effective coupling $f$ of Eq.(\ref{e37}), see Eq.(\ref{e44}). The solid
line shows the total $f_S$, while the dot--long dashed, long--short dashed and
long dashed curves show the contributions from the first, second and fourth
term in Eq.(\ref{e44}). The short dashed curve shows the contribution from the
third term in Eq.(\ref{e44}), where we have also included $\tilde{b}_1$
exchange as described in the text. The dot--short dashed and dotted curves
show the total squark exchange contribution to $f$ and the total result for
$f$, including Higgs exchange, respectively.

\item 
Contours of constant counting rate [in events/(kg$\cdot$day)] in a Germanium
detector. The region between the dotted lines at small $|M_2|$ is excluded by
sparticle searches, while the regions below the dotted curves at small
$m_{\tilde{q}}$ and large $|M_2|$ are excluded because here the LSP is
charged. In the hatched regions labelled $\Omega h^2>1$ relic
neutralinos overclose the universe, while in the shaded regions the relic
density is too small to make the LSP an interesting DM candidate. Figs a) and
c) are for a pure ${^76}$Ge detector with $\tan\beta=2$ and 8, respectively,
while b) is for a pure ${^73}$Ge detector and $\tan\beta=2$.

\item 
Contours of constant counting rate in a ${^76}$Ge detector as predicted in a
minimal supergravity model with radiative symmetry breaking. The notation is
as in Fig.5.

\item 
The $\tan\beta$ dependence of the counting rate as predicted in minimal
supergravity. Different curves correspond to different values of the common
scalar mass $m_0$ at scale $M_X$, as indicated. The curves are terminated at
high $\tan\beta$ where the rescaled relic density $\Omega h^2$ falls below
0.05. For $m_0$=300 GeV one has $\Omega h^2>1$ for $\tan\beta\leq$ 14.

\end{enumerate}
\end{document}